\documentstyle[seceq,mbf,epsf,wrapft]{ptptex}

\def\GA{\raise2.5pt\hbox{$>$}\kern-8pt\lower2.5pt\hbox{$\sim$}}
\def\LA{\raise2.5pt\hbox{$<$}\kern-8pt\lower2.5pt\hbox{$\sim$}}

\title{
Hund's-Rule Coupling Effect in Itinerant Ferromagnetism
}

\author{
Takuya {\sc Okabe}
}

\inst{
Department of Physics, Kyoto University, Kyoto 606-01
}

\recdate{
November 30, 1996}

\abst{
We present a general model
which includes the ferromagnetic Kondo lattice and the Hubbard model
as special cases.
The stability of the ferromagnetic state is 
investigated variationally.
We discuss the mechanism of 
ferromagnetism in metallic nickel,
emphasizing the importance of orbital degeneracy
and the effect of the Hund's-rule coupling.
}

\begin{document}

\maketitle

\section{Introduction}
\label{IM}

The study of itinerant ferromagnetism 
is one of the most difficult problems in solid-state theory.
In the 1960's, this problem was studied 
for the single-band Hubbard model
by Kanamori,~\cite{rf:Kanamori} Gutzwiller~\cite{rf:Gutzwiller}
and Hubbard.~\cite{rf:Hubbard}\
In the Hubbard model,
it is now widely known that the ferromagnetic ground state 
is realized only in exceptional situations.
An example is the Nagaoka ferromagnetic state,~\cite{rf:Nagaoka}\
which is shown to be the exact ground state 
when a single carrier is doped in 
the half-filled $U=\infty$ Hubbard model.
The other cases were discussed by
Tasaki and Mielke,~\cite{rf:Tasaki} who investigated the model
in which the relevant band is flat or almost flat.
Furthermore, 
Penc et al.~\cite{rf:Penc} discussed ferromagnetism
in such a way that their formulation includes flat-band limit as
a special case.
Our theory presented here is also aimed at generalizing the model for
itinerant ferromagnetism so as to include the Nagaoka state
in the Hubbard model.
Our generalization is
made by including the effect of the Hund's-rule coupling,
rather than by considering the specific lattice 
structure.~\cite{rf:Penc,rf:comment0}

In a pioneering work on the electron correlation effect,
Kanamori~\cite{rf:Kanamori}\ discussed ferromagnetism of Ni.
Although he started with degenerate $3d$ bands,
he finally reduced the problem
to that of a single band model.
This is partly due to overestimation of the interaction energy,
which was regarded as being roughly twice
as large as the band width.
With respect to this point, 
it is now commonly accepted that the interaction energy and the
band width in Ni are
of the same order of magnitude.~\cite{rf:Antonides,rf:Fuggle,rf:Bennett}\
Therefore, it is necessary to 
reconsider the effect of the Hund's-rule coupling.

In ferromagnetic nickel,  
satellite structure was observed
at 6 eV below the Fermi energy.~\cite{rf:XPS1,rf:XPS2,rf:UPS1,rf:UPS2}\
This structure is interpreted as 
a two-hole bound state.~\cite{rf:Penn,rf:Treglia}\
To reproduce the structure,
one must take into account
electron-electron and electron-hole multiple scattering
effects. 
For this purpose,
formal treatments have been given 
by many authors,~\cite{rf:Edwards1,rf:Brandt,rf:Roth,rf:Hertz,rf:Edwards2}\
and the resulting mathematical expressions
are all in similar forms.
In conformity with these results,
several works
have been presented to explain 
the structure.~\cite{rf:Liebsch1,rf:Liebsch2,rf:Igarashi1} \
The conclusion
is that it can be explained in the single-band model
if one assumes the Hubbard repulsion
$U\LA W\sim$4 eV.~\cite{rf:Penn,rf:Treglia,rf:Igarashi1}\
However, it then becomes difficult
to maintain the complete ferromagnetic ground state
for such a small value of $U$.~\cite{rf:Edwards2,rf:Igarashi1}\
Therefore the effect of degenerate $d$-orbitals
and the Hund's-rule coupling should be considered.

We investigated itinerant ferromagnetism in $3d$ transition metal
systems in
the generalized Gutzwiller approximation,~\cite{rf:Okabe1}
and concluded that 
the atomic correlation effect is quite effective 
in realizing ferromagnetism in Ni.
In this paper, more precise discussion
on the local stability condition of the
complete ferromagnetic state is given.
We restrict our argument to absolute zero temperature.
Not only the longitudinal but also 
transverse component of the Hund's-rule coupling 
is taken into account.
We reproduce the theoretical expressions
derived for the Hubbard model
by considering the multiple 
magnon scattering effect~\cite{rf:Edwards1,rf:Roth,rf:Hertz,rf:Edwards2}\
as a special case of our general result.

In the next section, we present a general model 
to treat itinerant ferromagnetism.
The model includes
the ferromagnetic Kondo lattice model,
which was treated in our previous paper.~\cite{rf:Okabe2}
In \S\S\ref{SW} and \ref{IPE},
the formulae required in the following sections are derived.
As an example, we discuss a two-band model,
which consists of a pair of Hubbard models coupled by the Hund's-rule 
coupling (\S\ref{TbM}).
Then a simple theory for metallic nickel is given in \S\ref{FiNM}.
It is concluded that the Hund's-rule coupling effect is indispensable.
Conclusion is given in the last section, \S\ref{DC}.

\section{General model for the itinerant ferromagnet}
\label{GMfIF}
As a general model for an itinerant ferromagnet,
we begin with  the Hamiltonian
\begin{eqnarray}
H&=&T+V, \\ 
T&=&-\sum_{i,j,\mu,\nu,\sigma
}t_{ij}^{\mu \nu}
c^{\dagger}_{i\mu\sigma}c_{j\nu\sigma}, \nonumber \\
V&=&U\sum_{i,\mu}\hat{n}_{i\mu\uparrow}\hat{n}_{i\mu\downarrow}
+U^{\prime}\sum_{i,<\mu\neq\nu>}  \hat{n}_{i\mu}\hat{n}_{i\nu}
-J\sum_{i,<\mu\neq\nu>} {\mbf s}_{i\mu} \cdot{\mbf s}_{i\nu},
\end{eqnarray}
where 
\begin{eqnarray}
\hat{n}_{i\mu\sigma}&\equiv&c^{\dagger}_{i\mu\sigma}c_{i\mu\sigma},
\nonumber\\
\hat{n}_{i\mu}&\equiv&\sum_{\sigma}\hat{n}_{i\mu\sigma},
\nonumber\\
{\mbf s}_{i\mu}&\equiv&\frac{1}{2}
\sum_{\sigma,\sigma'}c^{\dagger}_{i\mu\sigma}
{\mbf \sigma}_{\sigma,\sigma'}c_{i\mu\sigma'},
\end{eqnarray}
with the Pauli matrices ${\mbf \sigma}$.
The subscripts $i$, $j$ and $\mu$, $\nu$
designate lattice sites and localized orbitals, respectively.
The summation with $<\mu\neq\nu>$ denotes that
the sum is taken over pairs of orbitals $(\mu,\nu)$.

For simplicity, we assume 
that each component of degenerate bands
is labeled by the index $\mu$, or
$t^{\mu\nu}_{ij}=t^{\mu}_{ij}\delta_{\mu\nu}$.
Then we have
\begin{equation}
T=\sum_{k,\mu,\sigma}\varepsilon_{k\mu}
c^{\dagger}_{k\mu\sigma}c_{k\mu\sigma},
\label{HamT}
\end{equation}
where
\begin{equation}
\varepsilon_{k\mu}=-\sum_{j}t^\mu_{ij}e^{ik(r_j-r_i)}.
\end{equation}

If we further assume
a tight-binding band, or
$t^{\mu}_{ij}=t_\mu$
for the nearest-neighbor pair $i,j$ and zero 
for any other pairs $i\neq j$,
we have
\begin{equation}
\varepsilon_{k\mu}=-t_\mu\sum_{\delta} e^{ik\delta}-t_{ii}^\mu,
\end{equation}
where the summation is taken over the nearest neighbor vector $\delta$.
The second term is a constant shift of energy.
In the following sections, we 
assume $t_{ii}^\mu=0$ and use the $\mu$-dependent
Fermi energy $\varepsilon_{f\mu}$.
Then, 
\begin{equation}
\sum_k \varepsilon_{k\mu}=0.
\end{equation}
This assumption does not cause any difficulty since
we are prohibiting the inter-band transfer of particles
by the assumption made above (\ref{HamT}).

The interaction part is rewritten as
\begin{eqnarray}
V&=&U\sum_{i,\mu}\hat{n}_{i\mu\uparrow}\hat{n}_{i\mu\downarrow} \nonumber\\
&+&(U^{\prime}-\frac{J}{4})\sum_{i,<\mu\neq\nu>,\sigma}
\hat{n}_{i\mu\sigma}\hat{n}_{i\nu\sigma}
+(U^{\prime}+\frac{J}{4})\sum_{i,\mu\neq\nu}
\hat{n}_{i\mu\uparrow}\hat{n}_{i\nu\downarrow} \nonumber\\
&-&\frac{J}{2}\sum_{i,<\mu\neq\nu>} 
(s^{+}_{i\mu}s_{i\nu}^{-}+ s^{-}_{i\mu}s_{i\nu}^{+}),
\end{eqnarray}
where
\begin{equation}
s^{\pm}_{i\mu}\equiv({\mbf s}_{i\mu})_x \pm i({\mbf s}_{i\mu})_y,
\end{equation}
or, explicitly,
\begin{eqnarray}
s^+_{i\mu}&=&c^\dagger_{i\mu\uparrow}c_{i\mu\downarrow},\nonumber\\
s^-_{i\mu}&=&c^\dagger_{i\mu\downarrow}c_{i\mu\uparrow}.
\end{eqnarray}
As the ferromagnetic ground state itself is well described
by band theory,~\cite{rf:Morruzi}
it is safely assumed that
the electron correlation effect is irrelevant
in the ferromagnetic state.
In fact, this is true
in the single-band Hubbard model, where
the Pauli principle makes interaction irrelevant.
If the complete ferromagnetic ground state is uncorrelated, 
an analytical treatment to estimate 
the elementary excitation energy becomes feasible.
Therefore, for this purpose we assume $U^{\prime}=J/4$.
Then,
\begin{eqnarray}
\label{HamV}
V&=&U\sum_{i,\mu}\hat{n}_{i\mu\uparrow}\hat{n}_{i\mu\downarrow} \nonumber\\
&-&\frac{J}{2}\sum_{i,<\mu\neq\nu>}\left(
s^{+}_{i\mu}s_{i\nu}^{-}+ s^{-}_{i\mu}s_{i\nu}^{+}
-(\hat{n}_{i\mu\uparrow}\hat{n}_{i\nu\downarrow}+
\hat{n}_{i\mu\downarrow}\hat{n}_{i\nu\uparrow})\right). 
\end{eqnarray}

Finally we have two positive parameters $U$ and $J$
to describe the interaction effect.
In our model,
the term energy of the singlet state formed 
by two electrons in the same (different) orbital is $U$ ($J$),
while that of the triplet state is zero.~\cite{rf:comment}\
For example, for metallic nickel,
the relative order is 
$J\LA U \LA W\sim 4$ eV, where
$W$ is the band width.
In the following sections, we investigate
the model Hamiltonian which 
 consists of (\ref{HamT}) and (\ref{HamV}), i.e., 
\begin{eqnarray}
H&=&T+V\nonumber\\
&=&\sum_{k,\mu,\sigma}\varepsilon_{k\mu}
c^{\dagger}_{k\mu\sigma}c_{k\mu\sigma} 
+U\sum_{i,\mu}\hat{n}_{i\mu\uparrow}\hat{n}_{i\mu\downarrow} \nonumber\\
&-&\frac{J}{2}\sum_{i,<\mu\neq\nu>}\left(
s^{+}_{i\mu}s_{i\nu}^{-}+ s^{-}_{i\mu}s_{i\nu}^{+}
-(\hat{n}_{i\mu\uparrow}\hat{n}_{i\nu\downarrow}+
\hat{n}_{i\mu\downarrow}\hat{n}_{i\nu\uparrow})\right). 
\label{HamTV}
\end{eqnarray}
The complete ferromagnetic ground state $|{\rm F}\rangle$ is given by
\begin{equation}
|{\rm F}\rangle\equiv\prod_{\varepsilon_{k\mu}<\varepsilon_{f\mu}}
c^{\dagger}_{k\mu \uparrow}|0\rangle,
\label{F}
\end{equation}
where $|0\rangle$ denotes the vacuum
and $\varepsilon_{f\mu}$ is the Fermi energy.

The above model contains several models 
which have been intensively discussed in the field of 
the itinerant ferromagnetism.
For example, 
we recover the single-band Hubbard Hamiltonian
by neglecting orbital degeneracy.
On the other hand, we reproduce the ferromagnetic Kondo lattice model
with the localized spin $S_f=D/2$.
For this, we may
assume $n_\mu=1$ for bands with $\mu=1,2,3,\cdots, D$,
and $n_{D+1}\equiv n$ for the conduction band
($0\le n \le 1$).
The completely filled band ($n_\mu=1$) represents 
the Mott insulator and acts as an array of localized spins,
$s_\mu=1/2$.
These localized spins
are coupled to form $S_f=D/2$ by
the on-site Hund's rule.
To describe metallic nickel, one may assume 
triply-degenerate bands with $n_\mu=0.2$ for $\mu=1,2$ and 3
(\S\ref{FiNM}).

\section{Spin wave}
\label{SW}
In this and the next section we derive 
formulae required in the following sections.
These are generalization of our previous results.~\cite{rf:Okabe2}\

\subsection{Approximation I}
\label{SWAI}
First we consider the trial state for the spin-wave-excited state
\begin{equation}
\Psi_{0q}=
\Lambda^{-1/2}\sum_{i}e^{iqr_{i}}S_{i}^{-}|{\rm F}\rangle,
\label{Psi0}
\end{equation}
where
\begin{eqnarray}
S^-_{i}\equiv\sum_{\mu}s^-_{i\mu},
\label{totspin}
\end{eqnarray}
and $\Lambda$
is the total number of lattice sites.
Description with this trial state corresponds
to the random phase approximation (RPA) in the strong 
coupling limit.

The energy expectation value of 
(\ref{Psi0}) gives 
the magnon dispersion,~\cite{rf:Okabe2}
\begin{equation}
\label{omega0}
\omega_{0q}=\frac{1}{2S}
\cdot \frac{1}{\Lambda}\sum_{k,\mu}n_{k\mu}
(\varepsilon_{k+q \mu}-\varepsilon_{k\mu}),
\end{equation}
where
\begin{eqnarray}
n_{k\mu}&=&\langle {\rm F}|c^{\dagger}_{k\mu\uparrow}c_{k\mu\uparrow}
|{\rm F}\rangle, \nonumber \\
n_{\mu}&=&\sum_k n_{k\mu}, \nonumber\\ 
2S&\equiv &\sum_{\mu}n_{\mu}. 
\end{eqnarray}
In these expressions,
$n_{k\mu}$ is the step function,
$n_{\mu}$ is a carrier density in the band $\mu$,
and $S$ represents the total spontaneous magnetization.

In the long wavelength limit, we have
\begin{equation}
\omega_{0q}=D_{0} q^2\equiv
\sum_{\mu}D_{0\mu} q^2, \qquad (q\rightarrow 0)
\end{equation}
where
\begin{equation}
\label{d0mu}
D_{0\mu}=\frac{|\epsilon_{g\mu}|}{2Sz},
\end{equation}
and
\begin{equation}
\label{eg}
\epsilon_{g\mu}\equiv\frac{1}{\Lambda}\sum_{k}
n_{k\mu}\varepsilon_{k\mu}.
\end{equation}
Here we have assumed a tight-binding band with the coordination
number $z$.
The negative quantity $\epsilon_{g\mu}$
is the total kinetic energy per site of the band $\mu$. 
Note that energy of the ground state $|{\rm F}\rangle$
is given by 
\begin{equation}
E_{\rm g}=\Lambda\sum_\mu \epsilon_{g\mu} \quad< 0.
\end{equation}
Thus the magnon stiffness constant $D_0$ is 
proportional to $|E_{\rm g}|$:
\begin{equation}
\label{d0}
D_{0}=
\frac{|E_{\rm g}|}{2Sz\Lambda}.
\end{equation}

\subsection{Approximation II}
\label{SWAII}
It is well known that the RPA overestimates the stability 
of a ferromagnetic state.
To improve upon (\ref{Psi0}),
we postulate the state~\cite{rf:Okabe2}
\begin{equation}
\Psi_{q}=\Lambda^{-1/2}\sum_{i}e^{iqr_{i}}S_{i}^{-}|i\rangle,
\label{Psiq}
\end{equation}
where 
\begin{equation}
\label{i}
|i\rangle =
\Lambda^{-1/2}\sum_{j,\mu}(f_\mu^+(r_j-r_i)c^{\dagger}_{i\mu}c_{j\mu}
+f_\mu^-(r_j-r_i)c_{i\mu}c^{\dagger}_{j\mu})|{\rm F}\rangle, \\ 
\end{equation}
where we expressed $c_{i\mu\uparrow}$ simply as $c_{i\mu}$.
The state $\Psi_{q}$ is constructed so as not to 
suffer any energy loss by $U$ nor $J$.
Therefore it is used as a trial state for the case $U=J=\infty$.
The functions $f_\mu^\pm(r_j-r_i)$ are determined variationally.
Note that (\ref{Psi0}) is recovered by assuming
$f_\mu^\pm(r_j-r_i)=\delta_{r_j-r_i}$.

We introduce the magnon creation operator 
$S^{\dagger}_q$ defined by 
\begin{equation}
\Psi_{q}\equiv S^{\dagger}_q|{\rm F}\rangle.
\end{equation}
Then the energy of the variational state (\ref{Psiq})
is given by 
\begin{equation}
\omega_q=\frac{\langle {\rm F}|S_q [H,S^{\dagger}_q]|{\rm F}\rangle}
{\langle{\rm F}|S_q S^{\dagger}_q|{\rm F}\rangle}.
\label{omega_q}
\end{equation}
After some calculations, we obtain
\begin{eqnarray}
\label{omegaden}
\langle{\rm F}|
S_q S^{\dagger}_q|{\rm F}\rangle&=&\frac{1}{\Lambda}
\sum_{k\mu} \Delta_{k\mu}^0
f_{k\mu}^* f_{k\mu}
+\frac{1}{\Lambda^2}\sum_{k,p,\mu}
\Gamma_{kp\mu}^0f_{k\mu}^*f_{p\mu}, \\
\label{omeganum}
\langle{\rm F}|S_q[H,S^{\dagger}_q]|{\rm F}\rangle &=& 
\frac{1}{\Lambda}\sum_{k,\mu} \Delta_{kq\mu}
f_{k\mu}^*f_{k\mu}+\frac{1}{\Lambda^2}
\sum_{k,p,\mu}\Gamma_{kpq\mu}f_{k\mu}^*f_{p\mu},
\end{eqnarray}
where $f_k=n_kf_k+(1-n_k)f_k\equiv f^+_k+f^-_{k}$ is 
a sum of the Fourier transform of $f_\mu^\pm(r_j-r_i)$.
Quantities appearing in the above expressions are calculated
as follows;
\begin{eqnarray}
\Delta_{k\mu}^0 
&=& (2S-n_\mu) n_\mu(1-n_{k\mu})+(2S-n_\mu +1)(1-n_\mu)n_{k\mu}, \\
\Gamma_{kp\mu}^0 &=& (2S-n_\mu) (1-n_{k\mu})
(1-n_{p\mu})+(2S-n_\mu +1)n_{k\mu}n_{p\mu}, \\
\Delta_{kq\mu} &=& 
(2S-n_\mu) \left(n_\mu\varepsilon_{k\mu} -\frac{1}{\Lambda}\sum_{k^{\prime}}
n_{k^{\prime}\mu}\varepsilon_{k^{\prime}\mu}
\right)(1-n_{k\mu})\nonumber\\
&+&(2S-n_\mu +1)\left(\frac{1}{\Lambda}\sum_{k^{\prime}}
(1-n_{k^{\prime}\mu})\varepsilon_{k^{\prime}\mu}
-(1-n_\mu)\varepsilon_{k\mu}\right)n_{k\mu}
 \nonumber\\
&+&\left((1-n_\mu)^2\varepsilon_{k+q\mu}-\frac{1}{\Lambda^{2}}
\sum_{k',k''}n_{k'\mu}n_{k''\mu}
\varepsilon_{k'-k''-k-q\mu}
\right)n_{k\mu}, \\
\Gamma_{kpq\mu} &=& 
2(2S-n_\mu)\varepsilon_{k\mu}
\bigl(n_{k\mu}(1-n_{p\mu})-(1-n_{k\mu})n_{p\mu}\bigr) \\
&+&\left(\frac{1}{\Lambda}\sum_{k'}n_{k'\mu}\left(
\varepsilon_{k'+q\mu}+\varepsilon_{k'
-k-p-q\mu}\right)
+(1-n_\mu)(\varepsilon_{p+q\mu}+\varepsilon_{k+q\mu})\right)
n_{k\mu}n_{p\mu}. \nonumber
\end{eqnarray}
If we assume 
$t^{\mu}_{ij}=t_\mu$ 
for the nearest-neighbor pair $i,j$,
$\Delta_{kq\mu}$ and $\Gamma_{kpq\mu}$ are expressed
in terms of $\epsilon_{g\mu}$, (\ref{eg}),
as
\begin{eqnarray}
\Delta_{kq\mu} &=&(2S-n_\mu) \left(n_\mu\varepsilon_{k\mu} 
+|\epsilon_{g\mu}|
\right)(1-n_{k\mu})\nonumber\\
&+&(2S-n_\mu +1)\bigl(|\epsilon_{g\mu}|
-(1-n_\mu)\varepsilon_{k\mu}\bigr)n_{k\mu}
 \nonumber\\
&+&\left((1-n_\mu)^2-(\frac{\epsilon_{g\mu}}{zt_\mu})^2\right)
\varepsilon_{k+q\mu}n_{k\mu}, \\
\Gamma_{kpq\mu} &=&
2(2S-n_\mu)\varepsilon_{k\mu}
\bigl(n_{k\mu}(1-n_{p\mu})-(1-n_{k\mu})n_{p\mu}\bigr) \nonumber\\
&+&\left(\frac{|\epsilon_{g\mu}|}{zt_\mu}\left(\varepsilon_{q\mu}+\varepsilon_{k+p+q \mu}\right)
+(1-n_\mu)(\varepsilon_{p+q\mu}+\varepsilon_{k+q\mu})\right)n_{k\mu}n_{p\mu}.
\end{eqnarray}

These are generalization of our previous results for the
ferromagnetic Kondo lattice model.~\cite{rf:Okabe2}\
Their correspondence becomes clear by introducing
the `localized-spin' $S_f$,
\begin{equation}
2S_f\equiv 2S-n_\mu.
\end{equation}
The quantity $S_f$ represents the total spin corresponding to 
all bands other than that containing the particle under consideration,
i.e., all bands except for $\mu$.
Generally, $S_f$  depends on 
the band index $\mu$ and can take an arbitrary value,
although it must be a half integer
in the ferromagnetic Kondo lattice model.~\cite{rf:Okabe2}\
We use this parameter $S_f$ 
in the following sections.
As noted previously,~\cite{rf:Okabe2}\
the case $S_f=0$ can be used to investigate
the $U=\infty$ Hubbard model.

We can estimate the spin-wave stiffness  $D$ analytically 
from (\ref{omega_q}).~\cite{rf:Okabe2}\
The result is
\begin{eqnarray}
\label{Dq2}
\omega_q&\equiv&Dq^2 \qquad (q\rightarrow 0)\nonumber\\
&=&\sum_\mu(D_{0\mu}-\Delta D_\mu)q^2,
\end{eqnarray}
where $D_{0\mu}$ is defined in (\ref{d0mu}),
and 
\begin{equation}
\label{deltaD}
\Delta D_\mu=\frac{1}{2S}
\frac{(1-n_\mu)^2I_\mu}{1+(|\epsilon_{g\mu}|/zt_\mu)(I_\mu/2t_\mu)},
\qquad(>0)
\end{equation}
\begin{equation}
\label{Imu}
I_\mu=\frac{1}{\Lambda}\sum_k \frac{v_{k\mu}^2}{\Delta_{k0\mu}}n_{k\mu},
\end{equation}
\begin{equation}
v_{k\mu}=\frac{\partial \varepsilon_{k\mu}}{\partial k}.
\end{equation}
These expressions for $D$, below (\ref{Dq2}),
will be used in the following sections.

\section{Individual-particle excitation}
\label{IPE}
To describe the individual-particle excitation
which has a spin component antiparallel to the spontaneous magnetization,
we introduce the following creation operator
for the quasiparticle in the band $\mu$:
\begin{equation}
\label{Ckmu}
C^{\dagger}_{k\mu\downarrow}=
\frac{1}{\Lambda}\sum_{i,j}e^{ikr_i}\left(h_{\rm U}
c_{i\mu\downarrow}^{\dagger}+
\sum_\nu h^\nu_{j-i\mu}s_{i\nu}^{-} c^{\dagger}_{j\mu} 
\right).
\end{equation}
The expression (\ref{Ckmu}) for 
$h_{\rm U}=0$ and $h^\nu_{j-i\mu}=
\delta_{ij}$
appears as a prefactor of $c_{j\mu}$ in $\Psi_q$.
In this case, the trial state does not suffer
energy loss due to $U$ nor $J$.
To include the effect of finite interactions,
we introduce a variational parameter $h_{\rm U}$
and assume
the dependence of $h^\nu_{j-i\mu}$ on $\mu-\nu$.
The dependence on $r_j-r_i$ is 
further assumed to afford a better variational description.
This corresponds to allowing an up-spin particle 
($c^{\dagger}_{j\mu}$ in (\ref{Ckmu}))
to be created at a site different than the site $i$  where 
a down-spin particle is created by $s_{i\nu}^{-}$.
As will be seen below,  the effect of 
multiple magnon scattering is taken into account
by this generalization.

The energy expectation value of (\ref{Ckmu})
is given by
\begin{equation}
\label{Ekmu}
E_{k\mu\downarrow}
=\frac{\langle {\rm F}|
C_{k\mu\downarrow}[H,C^{\dagger}_{k\mu\downarrow}]|{\rm F}\rangle}
{\langle {\rm F}|
C_{k\mu\downarrow}C^{\dagger}_{k\mu\downarrow}|{\rm F}\rangle},
\end{equation}
for which
we obtain
\begin{eqnarray}
\label{Cden}
\langle {\rm F}|C_{k\mu\downarrow}C^{\dagger}_{k\mu\downarrow}|{\rm F}\rangle 
&=& \frac{1}{\Lambda}\sum_{k_1,\nu}n_\nu 
|h^\nu_{k_1\mu}|^2
+\left|h_{\rm U}
+\frac{1}{\Lambda}\sum_{k_1}h^\mu_{k_1\mu}\right|^2,\\
\langle {\rm F}|C_{k\mu\downarrow}[H,C^{\dagger}_{k\mu\downarrow}]
|{\rm F}\rangle 
&=& \frac{1}{\Lambda}\sum_{k_1,\nu}
|h^\nu_{k_1\mu}|^2
\left( n_\nu \varepsilon_{k_1\mu}+
\frac{1}{\Lambda}\sum_{k_2}n_{k_2\nu}(\varepsilon_{k-k_1+k_2\nu}
-\varepsilon_{k_2\nu})
\right) 
\nonumber\\ 
&+&\left|h_{\rm U}+\frac{1}{\Lambda}\sum_{k_1}
h^\mu_{k_1\mu}\right|^2 \varepsilon_{k\mu} 
+
g_\mu |h_{\rm U}|^2 \nonumber\\
&+&\frac{J}{2}\sum_{\nu (\neq \mu)}n_\nu
\left( n_\mu \sum_{k_1}\bigl|h^\mu_{k_1\mu}-h^\nu_{k_1\mu}\bigr|^2
+\bigl|\sum_{k_1}(h^\mu_{k_1\mu}-h^\nu_{k_1\mu})\bigr|^2 \right),\nonumber\\
\label{Cnum}
\end{eqnarray}
where 
$h^\nu_{k\mu}$ is the Fourier transform of $h^\nu_{i\mu}$
multiplied by $1-n_{k\mu}$.
The parameter $g_\mu$ is defined by
\begin{equation}
g_\mu\equiv Un_\mu+JS_f,
\label{gmu}
\end{equation}
and represents the Hartree-Fock
exchange splitting of our model.

We remark that we must assume (i) $h_{\rm U}=0$ for $U=\infty$,
and (ii) $h^\nu_{k_1\mu}=h^\mu_{k_1\mu}$ and
$h_{\rm U}=0$ for $J=\infty$,
as noted above.
This is because the terms depending on $g_\mu$ and/or $J$ in (\ref{Cnum}),
which are positive definite,
should vanish 
in the strong coupling limit.

\subsection{Case $J\neq\infty$}
\label{IPECwJ}
As the above expressions are too general and complicated,
we show simpler expressions which are used 
in the following investigation.
To investigate the effect of a finite Hund's-rule coupling $J$,
we introduce a new variational parameter $h_{\rm J}$ by
\begin{equation}
h^\nu_{k\mu}=h_{\rm J}h^\mu_{k\mu}.
\label{hJ}
\end{equation}
The parameter $h_{\rm J}$ 
measures the effect of the Hund's-rule coupling.
In particular,
$h_{\rm J}=0$ for $J=0$, and $h_{\rm J}=1$ for $J=\infty$.

All terms but the last term of (\ref{Cnum})
are independent of the phase of the complex quantity $h_{\rm J}$.
Therefore,
it is always energetically unfavorable to
have $h^\nu_{k_1\mu}$ out of phase with $h^\mu_{k_1\mu}$.
Thus,
we may assume 
$h_{\rm J}$ as a real quantity, $0\le h^\nu_{k_1\mu} \le 1$.
This implies that the spins ${\mbf s}_{i\mu}$ at site $i$ 
should couple in phase with each other
by the Hund's-rule coupling.
For $h_{\rm J}=1$,
the spin lowering operator
$\sum_\nu h^\nu_{j-i\mu}s_{i\nu}^{-}$
in (\ref{Ckmu})
reduces to a quantity proportional to (\ref{totspin}).

After taking the functional derivative of (\ref{Ekmu})
with respect to $h_{\rm U}$ and $h^\mu_{k_1\mu}$
and eliminating these parameters,
one obtains the eigenequation 
for the excitation energy $E_{k\mu\downarrow}$,~\cite{rf:Okabe2}
\begin{eqnarray}
\label{eigeneqhj}
E_{k\mu\downarrow}-\varepsilon_{k\mu}&=&\Sigma_\mu(k,E_{k\mu\downarrow}),
\\ 
\label{eigeneqhj2}
\Sigma_\mu(k,E_{k\mu\downarrow})&=&
\dfrac{g_\mu\left(1-j_\mu
\tilde{\rho}\right)}{
1-\left(g_\mu+j_\mu
\right)
\tilde{\rho}}, \\   \nonumber
\end{eqnarray}
where 
\begin{equation}
\label{tilderho}
\tilde{\rho}\equiv
\frac{1}{\Lambda}
\sum_{k_1}\dfrac{1-n_{k_1\mu}}
{
(2S_f h_{\rm J}+n_\mu)(
E_{k\mu\downarrow}-\omega^\prime_{\mu k-k_1}-\varepsilon_{k_1\mu})
-n_\mu j_\mu
},
\end{equation}
\begin{equation}
\label{omegaprime}
\omega^\prime_{\mu q}\equiv
\frac{1}{2S_fh_{\rm J}+n_\mu}
\dfrac{1}{\Lambda}\sum_{k_2,\nu}
\bigl(h_{\rm J}+(1-h_{\rm J})\delta_{\mu\nu}\bigr)
n_{k_2\nu}(\varepsilon_{q+k_2\nu}
-\varepsilon_{k_2\nu}),
\end{equation}
and
\begin{equation}
j_\mu\equiv JS_f(1-h_{\rm J})^2.
\label{jmu}
\end{equation}

Noting the identity
\begin{equation}
\frac{1}{\Lambda}\sum_{k_2,\nu}n_{k_2\nu}
\bigl(h_{\rm J}+(1-h_{\rm J})\delta_{\mu\nu}\bigr)
=2S_fh_{\rm J}+n_\mu,
\end{equation}
and comparing $\omega^\prime_{\mu q}$ with (\ref{omega0}),
we observe that $\omega^\prime_{\mu q}$ is the generalization of the
magnon dispersion in the presence of the inter-band spin-spin coupling.
For the complete coupling $h_{\rm J}=1$,
(\ref{omega0}) is recovered,
while $h_{\rm J}=0$ gives
the single-band result for the band $\mu$.

Since further variation with respect to $h_{\rm J}$
leads to a complicated expression,
we leave $h_{\rm J}$ as a parameter.
Thus, the equation (\ref{eigeneqhj}) is to be solved 
for $E_{k\mu\downarrow}$ as a function of $h_{\rm J}$,
which in turn should be determined so as to minimize 
$E_{k\mu\downarrow}$.
The expressions following (\ref{eigeneqhj}) are
conclusions of this subsection.
We note that the derivation of (\ref{eigeneqhj}) 
from (\ref{Ckmu}) is exact.
While an assumption is made in (\ref{hJ}),
this does not conflict with the variational principle.
Introduction of the single parameter $h_{\rm J}$ 
to describe the Hund's-rule coupling is
largely for the sake of simplicity.

\subsection{Case ${\it h_{ J}=1}$}
\label{IPECwhJ1}
The case $h_{\rm J}=1$ is realized 
for $J=\infty$, where
the Hubbard repulsion $U$ is always ineffective.
This is because the strong Hund's-rule coupling 
forbids two particles to stay on the same site. 
However, there is a situation in which 
the assumption $h_{\rm J}=1$ and $h_{\rm U}\neq 0$
is adequate even for a finite interaction $J<\infty$.
We observed that this is in fact 
the case for the double exchange ferromagnet.~\cite{rf:Okabe2}\
By this assumption, the number of variational parameters
decreases by one, and the calculation becomes relatively easy.

In this case,
from the results of the previous subsection,
we obtain an eigenequation of the form
\begin{equation}
\label{E-eps-Sigma}
E_{k\mu\downarrow}-\varepsilon_{k\mu}-\Sigma_\mu(k,E_{k\mu\downarrow})=0,
\end{equation}
where
\begin{equation}
\Sigma_\mu(k,
\omega)=
\dfrac{g_\mu}{1-\Sigma_\mu^0(k,
\omega
)/g_\mu},
\label{E-eps-Sigma2}
\end{equation}
and
\begin{equation}
\label{Sigma0}
\Sigma_\mu^0(k,
\omega
)=
\frac{g_\mu^2}{2S}\frac{1}{\Lambda}\sum_{k_1}\frac{1-n_{k_1\mu}}
{
\omega
-\omega_{0k-k_1}-\varepsilon_{k_1\mu}}.
\label{E-eps-Sigma3}
\end{equation}
The magnon energy $\omega_{0q}$ is defined in 
(\ref{omega0}).

In these forms, 
it is possible to regard $\Sigma_\mu(k,\omega)$ as the self-energy
given as a result of the multiple scattering of 
quasiparticle $\varepsilon_{k_1\mu}$ off the spin wave $\omega_{0k-k_1}$.
In fact, a particular case of this result,
for the
single-band Hubbard model 
$2S=n$ and $g=Un$,
has already been obtained by many 
authors.~\cite{rf:Edwards1,rf:Roth,rf:Hertz,rf:Edwards2,rf:Liebsch2}
Although the formulae we derived have
the same forms as those
derived in the other methods,
it contains a meaning more significant than that of mere re-derivation.
Since our results are based on the variational principle,
we can draw a definite conclusion 
on the instability of the ferromagnetic state.

\section{Two-band model}
\label{TbM}
In this section we consider a simple case to show
general features of our model.
We investigate 
a special case of (\ref{HamTV}), i.e., 
a model comprising two bands
in a tight-binding square lattice structure.
Then the band dispersion is given by
$\varepsilon_{k\mu}=-2t(\cos(k_x)+\cos(k_y))$ ($\mu=1,2$, $t>0$),
and the band-width is $2zt$ with $z=4$.
This model includes
the Hubbard model and 
the $S_f=1/2$ ferromagnetic Kondo lattice model
as special cases.
Therefore we can treat them in a unified way.
Parameters used in the following discussion
are the interaction $U$, $J$, and the carrier density $n_\mu$
of the band $\mu$.
The purpose of this section is to investigate
the effect of the orbital degeneracy and the Hund's-rule coupling.

\subsection{Spin wave}
\begin{figure}[tb]
 \parbox{\halftext}{
\epsfxsize = 6.6 cm
\epsfbox{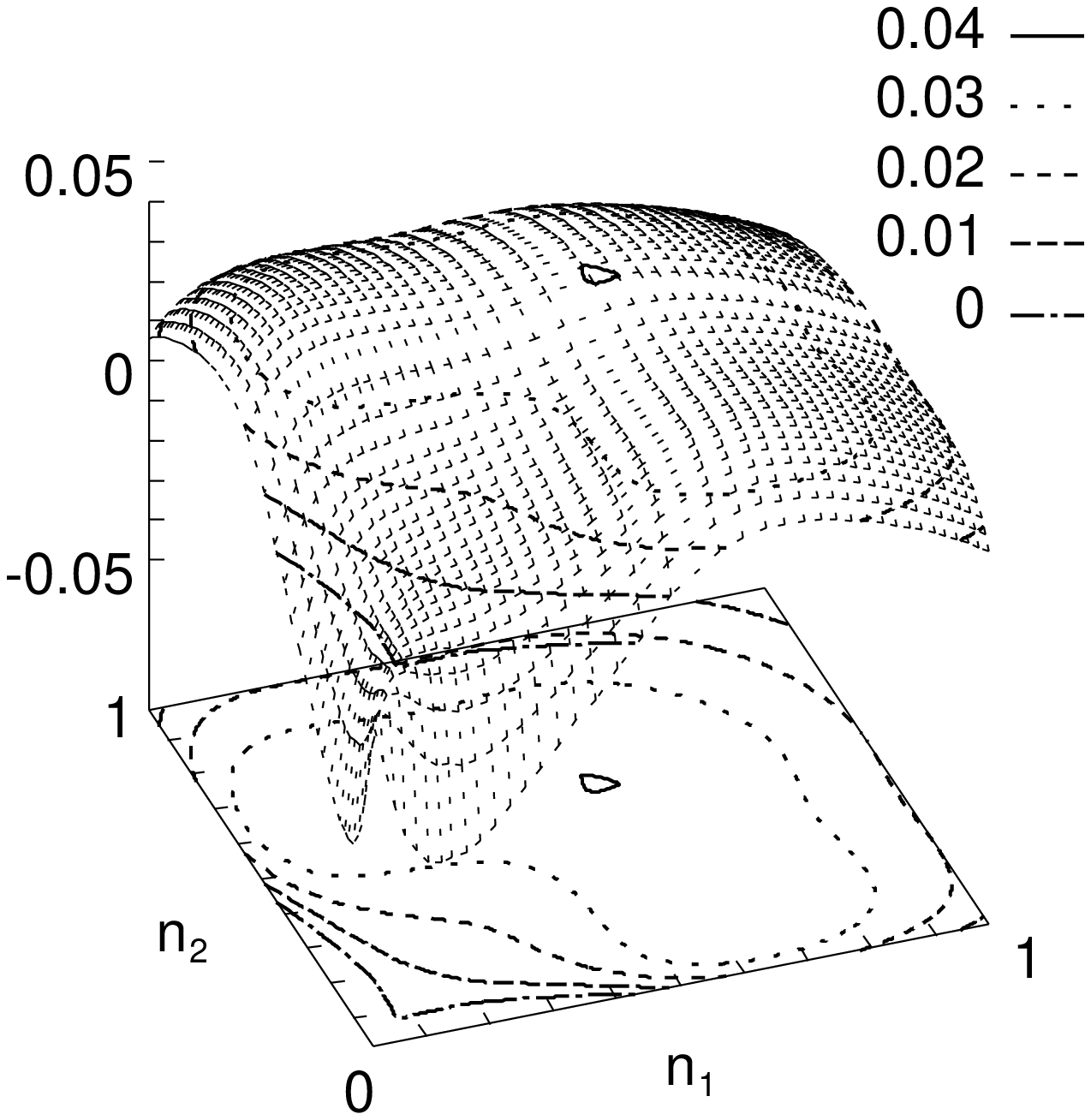}
}
 \hspace{6mm}
 \parbox{\halftext}{
\epsfxsize = 6.6 cm
\epsfbox{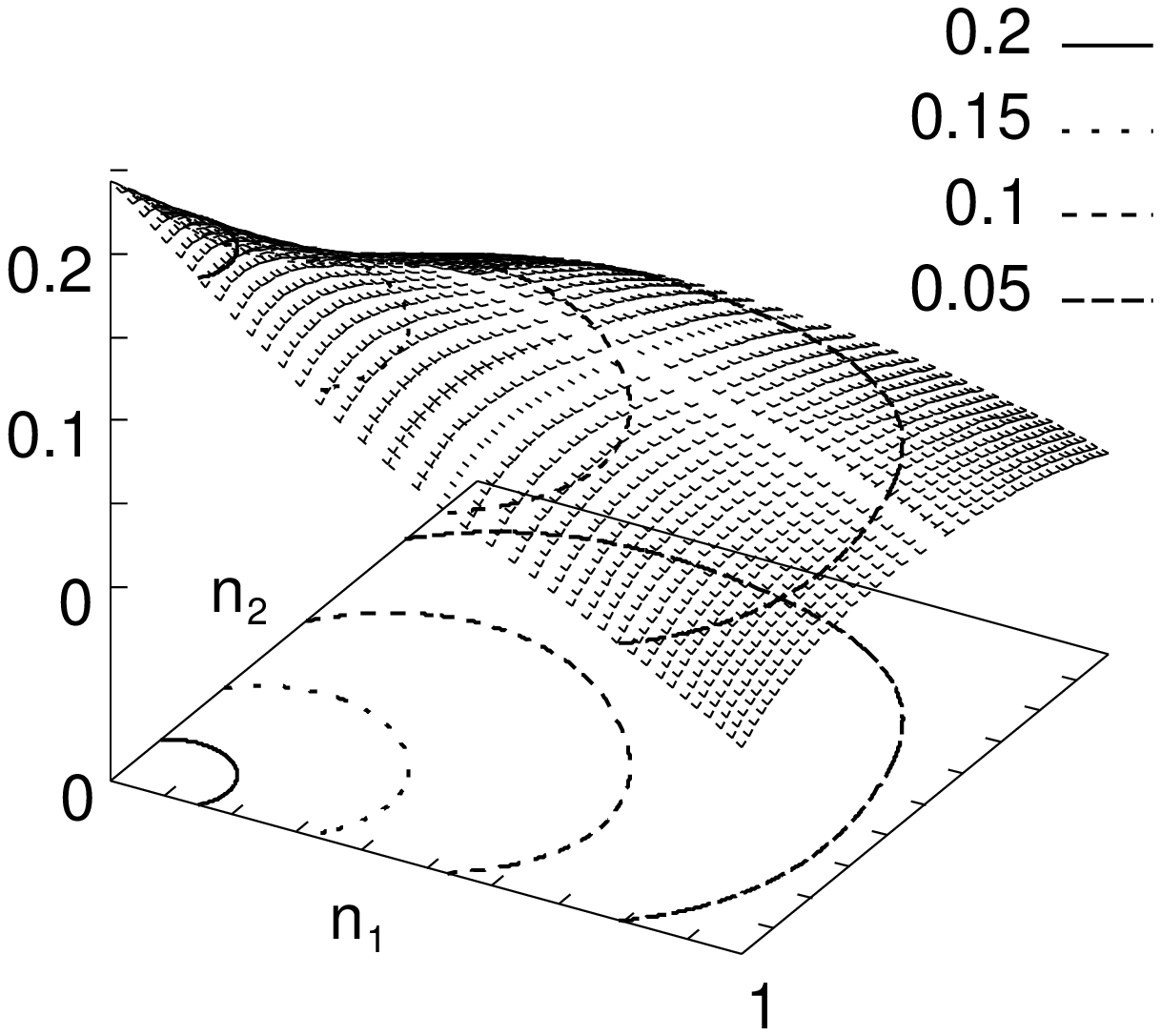}
}
\caption{Spin-wave stiffness constant $D(n_1,n_2)$ (left)
and $D_0(n_1,n_2)$ (right).
}
\label{fig:1}
\end{figure}

\begin{figure}[tb]
 \parbox{\halftext}{
\epsfxsize = 6 cm
\epsfbox{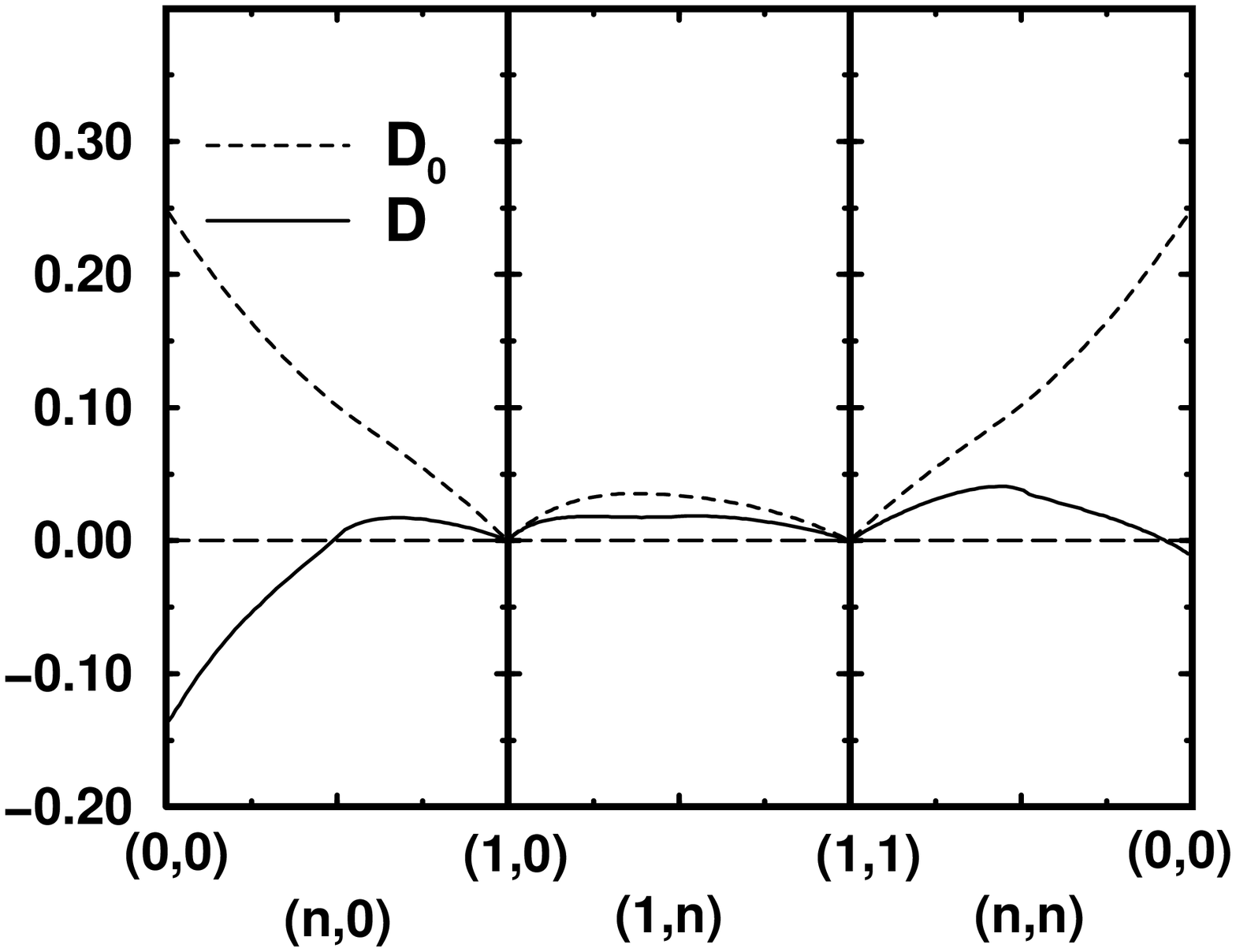}
\caption{
Spin-wave stiffness constant $D$
and $D_0$ as a function of $(n_1,n_2)$.
}
\label{fig:2}
}
 \hspace{6mm}
 \parbox{\halftext}{
\epsfxsize = 6 cm
\vspace{5mm}
\epsfbox{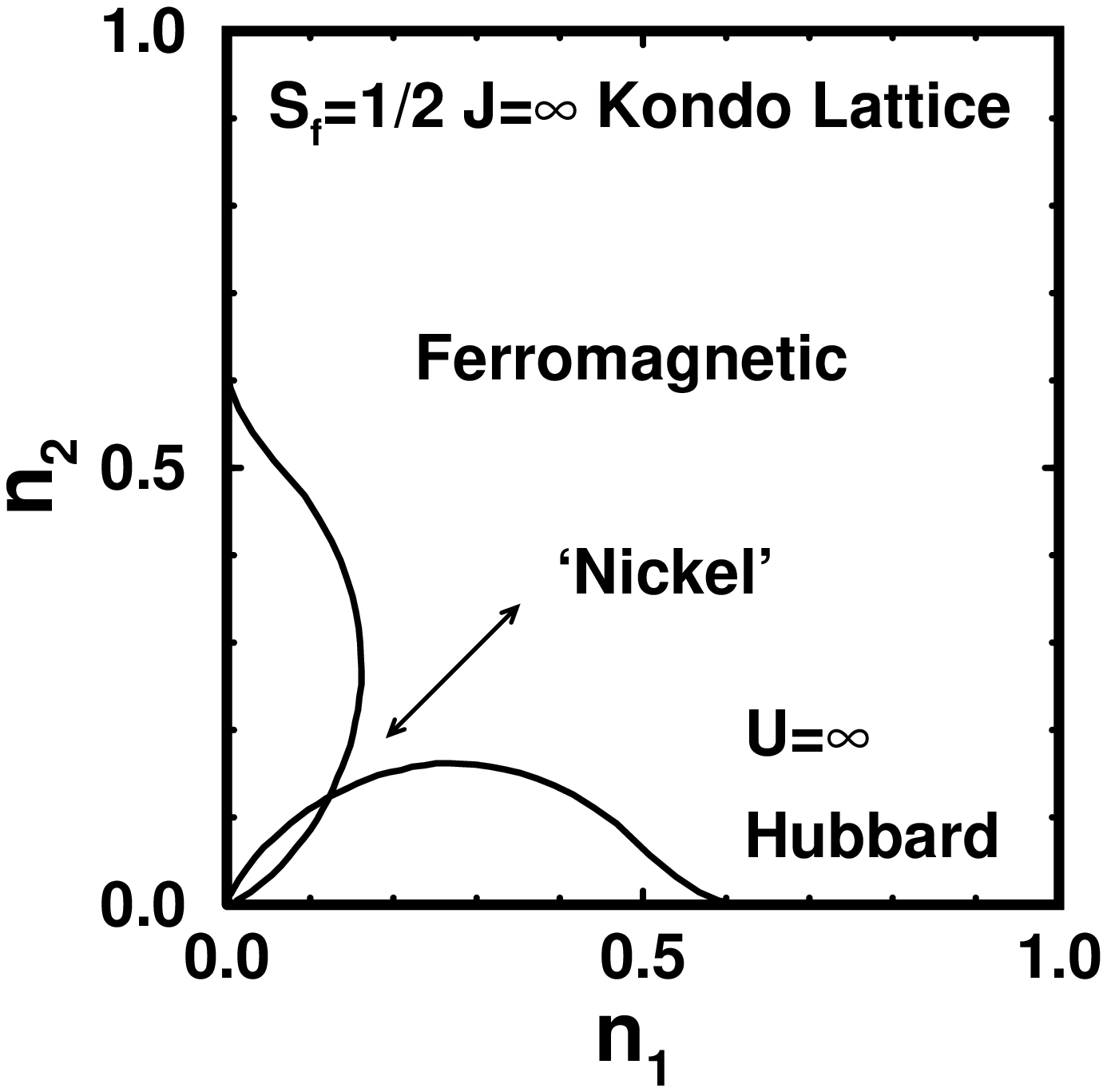}
\caption{
Phase boundary determined by the condition
$E_{k1\downarrow}=\varepsilon_{f}$
and $E_{k2\downarrow}=\varepsilon_{f}$ with $k=0$.
In the region enclosed by the curves,
the ferromagnetic state is absolutely unstable.
}
\label{fig:3}
}
\end{figure}

In this and the next subsection,
we consider the strong coupling limit $U=J$$=\infty$,
where the formulae given in \S\ref{SW} are of direct use.
In Fig.~\ref{fig:1}, we show the spin-wave stiffness constant
$D(n_1,n_2)$ as well as $D_0(n_1,n_2)$
as a function of $n_1$ and $n_2$,
where $D_0$ is given by (\ref{d0})
and $D$ is defined below (\ref{Dq2}).
In Fig.~\ref{fig:2}, they are
shown along the symmetry axes in the $n_1$-$n_2$ plane.

If the effect of band degeneracy 
is simply additive,
one should expect $D(n,n)$ $=$~$D(n,0)$ $=$~$D(0,n)$.
In fact,
this holds for the result $D_0$ in the RPA,
for which the  curve
determined by $D_0(n_1,n_2)$=const becomes convex in the direction
$(n_1,n_2)$ $=(1, 1)$ (Fig.~\ref{fig:1}).
However, this is not the case for the improved estimate $D(n_1,n_2)$;
the curve $D(n_1,n_2)$=0 is concave instead,
as shown in the dash-dotted
contour line on the left of Fig.~\ref{fig:1}.
This is due to a singularity present in $D(n_1,n_2)$
at $(n_1,n_2)$ $=(0,0)$.
Here we note that  $D(n_1,n_2)=0$ defines the threshold
for ferromagnetism.
In our treatment, the result $D<0$ represents the absolute instability 
of the ferromagnetic state.
From the displayed results,
we understand that a multi-band model favors ferromagnetism 
more than a single-band model
and that one should go beyond the RPA
to show this multi-band effect.
In the RPA,
we cannot prove instability at all,
since $D_0>0$  for any $(n_1,n_2)$ (see Fig.~\ref{fig:2}).
In particular, one should be careful in treating the low-density limit
owing to the presence of the singularity mentioned above.
This point is discussed later.

\label{TbMSW}
Next, we note that
the line along $(n,0)$ or $(0,n)$ ($0\le n\le 1$) corresponds to 
the $U$=$\infty$ Hubbard model and the line along $(n,1)$ or $(1,n)$ to
the ferromagnetic Kondo lattice model with $S_f$=1/2 and $J$=$\infty$.
These two cases were discussed in Ref.~\citen{rf:Okabe2}.
As a new result,
in the direction $(n,n)$, i.e., in the case $n_1=n_2$,
the singular behavior around (0,0)
makes the threshold value so small as $n_c\LA0.1$.
(See $D$ in Figs.~\ref{fig:1} and \ref{fig:2}).
We note that the ferromagnetic state is unstable
for $n<n_c$, while we may assume $n_{\rm Ni}\sim 0.2$ for Ni.
Thus one may simply consider that
the fact $n_{\rm Ni}>n_c$ 
is relevant to the itinerant ferromagnetism in Ni.
This point is addressed in the next subsection,
since the more stringent condition is obtained
by investigating the individual-particle excitation.

\subsection{Individual-particle excitation}
\label{TbMIPE}
On the stability of the ferromagnetic state, 
investigation of the individual-particle excitation
produces a 
more stringent condition
than that given by the spin-wave stiffness constant.~\cite{rf:Okabe2}\
To investigate the case $U=J=\infty$, we use 
the expressions given in 
\S\ref{IPECwhJ1}
The eigenenergy $E_{k\mu\downarrow}$,
regarded as a real quantity,
is calculated as a solution of 
(\ref{E-eps-Sigma}) $\sim$ 
(\ref{E-eps-Sigma3}).
We determined 
the critical boundary in the $n_1$-$n_2$ plane
using the condition $E_{k\mu\downarrow}=\varepsilon_{f\mu}$ ($\mu=1,2$),
where we set
$k=0$ for the bottom of the dispersion $\varepsilon_{k\mu}$.
The result is shown in Fig.~\ref{fig:3}.
The Fermi energy $\varepsilon_{f\mu}$ is fixed for a given $n_\mu$.
The ferromagnetic state is absolutely unstable
when $E_{k\mu\downarrow}<\varepsilon_{f\mu}$.

As mentioned,
threshold curves thus obtained give a more stringent condition
than that determined by the relation $D(n_1,n_2)=0$.
(Compare Fig.~\ref{fig:3} with the contour line $D=0$ 
of Fig.~\ref{fig:1}.)
In particular, in the direction $(n,n)$, 
the critical concentration is estimated as $n_c\sim0.12$.
In this respect, let us remark again that
the stiffness $D$ has a singularity,
as noted in the previous subsection.
This is due to 
the denominator $2S\rightarrow0$ as $n\rightarrow0$
in the expression for the stiffness constant $D$,
(\ref{deltaD}).
On the other hand, 
$D_0(n_1,n_2)$ goes to a finite value in this limit
$(n_1,n_2)$ $\rightarrow(0,0)$ (Fig.~\ref{fig:2}),
because
the numerator of (\ref{d0})
also vanishes in this limit.
Thus, the correction $\Delta D/D_0$
is largest in the vicinity of the origin, as seen in Figs.~\ref{fig:1}
and \ref{fig:2}.
On the other side, 
the equation (\ref{E-eps-Sigma}) for $E_{k\mu\downarrow}$
contains the spin-wave energy $\omega_0(q)$ in the 
denominator of $\Sigma_\mu^0(k,\omega)$, (\ref{Sigma0}).
Our calculation based on (\ref{Ckmu}) is exact,
so that our result gives
the exact lower bound for $n_c$.
Nonetheless, 
in a physical (i.e. non-rigorous) sense 
for improving the approximation,
one should replace
$\omega_0(q)\simeq D_0q^2$ in (\ref{Sigma0}) with
$\omega(q)\simeq Dq^2$.
In this replacement, a more stringent and realistic
condition will be obtained,
although it then loses variational significance.~\cite{rf:Okabe2}\
As a result, we expect that 
the critical boundary in the region where $D\ll D_0$ is
drastically modified as the approximation is improved.
For example, from Fig.~\ref{fig:2}
along the line $(1,n)$, i.e., for the double exchange ferromagnet
with $S_f$=1/2, the correction will be less prominent
than the case along $(n,0)$, the $U$=$\infty$ Hubbard model.
Around the origin,
$D$ differs considerably from $D_0$ (Figs.~\ref{fig:1} and \ref{fig:2}),
so that
the critical $n_c$ in a proper treatment will take
a larger value than $n_c\sim0.12$ of Fig.~\ref{fig:3}.
However, it is still legitimate to assert that
the doubly-degenerate-band model is more favorable 
to ferromagnetism than the single-band model,
although in the following section we stress that 
ferromagnetism of Ni cannot be explained 
only by the effect of orbital degeneracy,
but a peculiarity of the density of states is
required in addition (\S\ref{FiNM}).

In any case,
we note that the correction $\Delta D/D_0$ (\ref{deltaD})
is a quantity of order $\sim$~$O(1/S)$ for large $S$,
since $\Delta_{k0\mu}$ in $I_\mu$ (\ref{Imu})
is proportional to $S$.
Therefore, the RPA 
is quite reliable for models
with highly degenerate orbitals and/or a large spontaneous 
magnetization,
as is expected intuitively.

\subsection{The effect of the Hund's-rule coupling}
\label{TbMEoHrC}

In this subsection, 
we assume  $U$ to be infinite but $J$ to be finite in order
to consider the effect of the Hund's-rule coupling.
As in the above subsections, the model
consists of the two Hubbard models in the square lattice,
coupled to each other by the Hund's-rule coupling.

We are interested in the effect of the interaction $J$
which couples the two bands.
Previously it was concluded that 
the ferromagnetic state in the single-band Hubbard model
(the limit $J\rightarrow 0$ of the present model) is unstable
beyond some critical hole concentration $\delta_c\equiv1-n_c$,
while it is stable for an arbitrary concentration
in the $S_f$=1/2 Kondo lattice model (the limit $J\rightarrow\infty$).
On the other hand, as we saw in the previous subsections,
the model which  assumes $n_1=n_2$ 
is more stable than the single-band model,
although it would be less stable than the double exchange ferromagnet with 
$S_f$=1/2.
To interpolate these situations in terms of $J$,
the critical carrier density 
determined by the quasiparticle energy is given below
as a function of $J$.
To this end,
we use the formulae given in \S\ref{IPECwJ}

The eigenenergy $E_{k\mu\downarrow}$ is obtained as 
a solution of (\ref{eigeneqhj}) with 
\begin{equation}
\Sigma_\mu(k,E_{k\mu\downarrow})=
j_\mu-\frac{1}{\tilde{\rho}}, 
\end{equation}
since $g_\mu=\infty$ for $U=\infty$.
Here, $\tilde{\rho}$ and $j_\mu$ are
functions of $h_{\rm J}$ and $n_\mu$ (or $S_f$), 
and are defined in 
(\ref{tilderho}) and (\ref{jmu}), respectively.
As noted there, the variational parameter $h_{\rm J}$ has been left
as a parameter to be determined to minimize the energy.
However, in determining the phase boundary,
minimization on $h_{\rm J}$ is not necessary;
we only have to calculate 
$J$ to satisfy $E_{k\mu\downarrow}=\varepsilon_{f}$
for a fixed carrier density $n$
and for various values of $h_{\rm J}$.
The envelope formed by a set of curves for every $h_{\rm J}$
then gives the threshold $J$ as a function of $n$,
since the variational principle implies instability
in all cases if it is the case for any of $h_{\rm J}$.

\begin{figure}[t]
\parbox{\halftext}{
\vspace{3mm}
\epsfxsize = 6.6 cm
\epsfbox{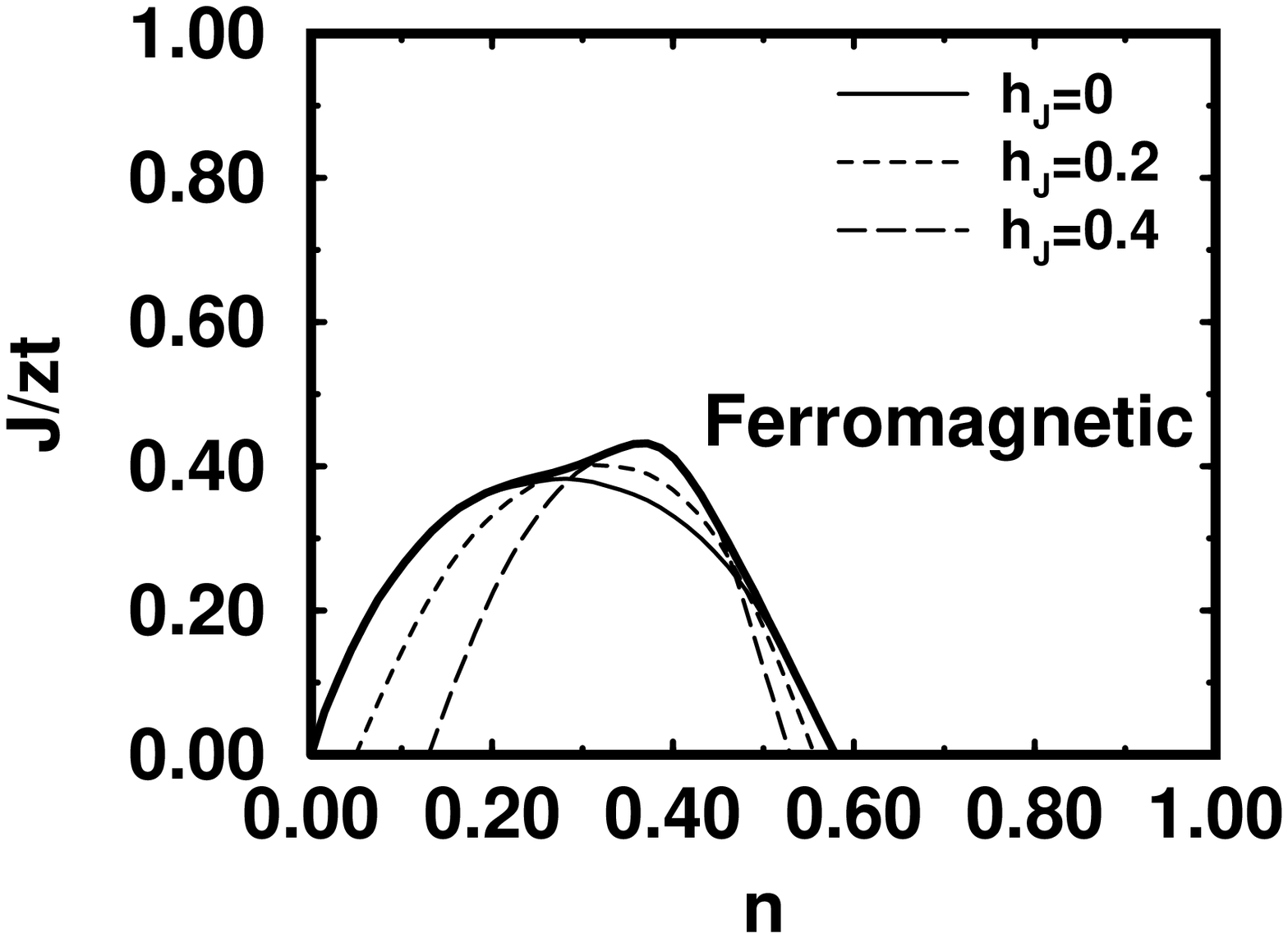}
\caption{
Thick line: Threshold for the stability of the
ferromagnetic state.
The model
consists of the $U=\infty$ Hubbard model
and  an array of localized spins $S_f=1/2$
coupled to each other by $J$.
}
\label{fig:4}}
\hspace{8mm}
\parbox{\halftext}{
\epsfxsize = 6.6 cm
\epsfbox{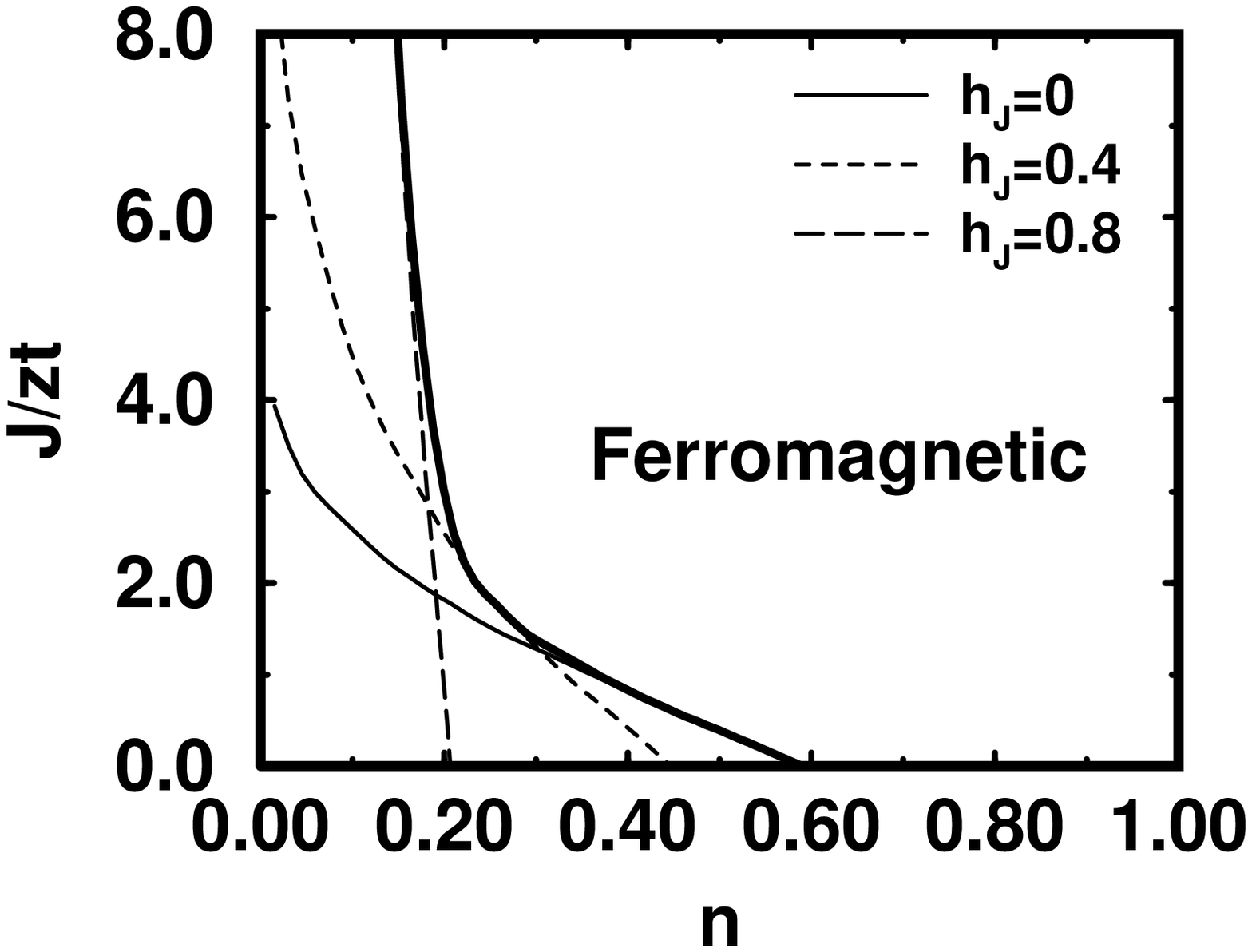}
\caption{
Thick line: Threshold for the stability of the
ferromagnetic state.
The model consists of
two equivalent $U=\infty$ Hubbard models
coupled by $J$.
}
\label{fig:5}
}
\end{figure}

We give results for the model (i) for which we set $n_1=1$ and $n_2=n$
(Fig.~\ref{fig:4}),
and for the model (ii) with $n_1=n_2=n$ 
(Fig.~\ref{fig:5}).
In the former, 
the single-band Hubbard model is
interpolated to the $S_f$=1/2 ferromagnetic Kondo lattice model.
In the latter,
it is interpolated as a function of $J$
to the case of two equivalent Hubbard models
strongly coupled via Hund's rule.

For the model (i) (see Fig.~\ref{fig:4})
the region in which the ferromagnetic state is stable is quite 
large.
In particular, the Hund's-rule coupling becomes 
effective even for $J$ as small as 0.5$zt$,
a quarter of the band-width $W=2zt$.
For large $J$
the ferromagnetic phase becomes stable 
for arbitrary $n$, as expected for the double exchange model.
The critical boundary (envelope, thick curve) in Fig.~\ref{fig:4}
is determined by curves for 
relatively small values $h_{\rm J}\LA0.4$,
as expected 
for the boundary in the weak-coupling region;
curves with larger $h_{\rm J}$ 
are completely surrounded by the envelope
and do not contribute to it.

For the model (ii) (see Fig.~\ref{fig:5})
the envelope strongly bends upward
around $J\sim2zt\equiv W$.
This implies  that
the Hund's-rule coupling becomes effective
when the coupling energy becomes comparable with the band width.
The critical hole concentration $\delta_c$ increases
from $\delta_c\sim 0.4$ for the $U$=$\infty$ Hubbard model
to $\delta_c\sim 0.9$
as $J$ increases from zero to
infinity.
This, and especially a steep increase in $\delta_c$ 
around $J\sim0$
clearly indicate the importance of the inter-band
spin-spin coupling.
Thus it is concluded that the stability condition
based on a single-band model becomes
more stringent than what should be obtained 
for realistic bands with orbital degeneracy.
The Hund's-rule coupling effect,
if not complete (or even if $h_{\rm J}\neq1$), 
can change the criterion for the ferromagnetic instability 
to a degree that it cannot be neglected 
in investigating real materials.
In the next section, we discuss
the itinerant ferromagnetism in Ni,
where the importance of the Hund's-rule coupling is again stressed.

\section{Ferromagnetism in nickel metal}
\label{FiNM}
Discussion given above has
been based on the expressions derived
variationally in \S\S\ref{SW} and \ref{IPE}.
Thus the results indicating instability of the ferromagnetic state
are exact.
However,
in discussing itinerant ferromagnetism of metallic nickel,
we prefer to give a physical argument 
since a model for Ni has to be simple enough to be tractable.
A model density-of-states that
we adopt below has some characteristic features
which are regarded to be relevant to
ferromagnetism in Ni.
We do not attempt to estimate 
the magnon dispersion, for example, by using our expression for $D$,
since we are not sure that 
the model given below is so quantitative as to reproduce
an appropriate estimate even for the spin wave spectrum.
Therefore, we regard the spin-wave stiffness constant $D$ 
as a given quantity and 
substitute the dispersion $\omega_q=Dq^2$
for $\omega^\prime_q$ in (\ref{tilderho}).
The model and parameters used below are set so as to 
mimic metallic nickel.

\subsection{Simple model simulating Ni}
To describe metallic nickel,
we adopt a simple model as a matter of convenience.
As one of important characteristic,
the density of states of Ni at the Fermi level $\rho_{f}$
takes a quite large value.
This is doubtless an important factor for ferromagnetism,
as indicated from the Stoner criterion $\rho_{f}U>1$.
Thus, as an example of models simulating Ni,
we adopt the model due to Edwards and Hertz,~\cite{rf:Edwards2}
i.e., with the isotropic dispersion $\varepsilon_k$ 
depending on $k$ quadratically:
\begin{equation}
\varepsilon_k=
\left\{
\begin{array}{lcl}
\displaystyle{\frac{k^2}{2m_1}}&,&
\qquad 0\le k\le k_1, \\
\noalign{\vskip0.2cm}
\displaystyle{\frac{k^2}{2m_2}-C}&,&
\qquad k_1\le k\le k_{\rm D}. \\
\end{array} \right.
\label{model}
\end{equation}
Here,
the constant $C$ is chosen 
to make $\varepsilon_k$ continuous at $k_1$.
The quantity $k_{\rm D}$ $=(6\pi^2)^{1/3}$ is a Debye-like cutoff
chosen so that the band contains one electron state per spin 
per orbital.
The parameters $m_1$ and $m_2$ 
are determined so as to meet the following conditions.
(i)\,For the Fermi wave-vector $k_{f}=k_1$, 
the total density  per band is $n=0.2$.
(ii)\,The bandwidth is 4 eV.
For the Fermi energy, we introduce a parameter
$\varepsilon_{1}(\equiv \varepsilon_{k_1})$
which characterizes a peculiarity of the density of states.
The parameter $k_1$ in (\ref{model}) is 
uniquely determined by the condition (i).
Then, $m_1$ is set by $\varepsilon_{k_1}=\varepsilon_{1}$
for a given $\varepsilon_{1}$.
Parameters $m_2$ and $C$ are determined 
to give the total band width 4 eV 
and $\varepsilon_{k_1-0}=\varepsilon_{k_1+0}$.
If we assume $\varepsilon_{1}=0.3$ eV to mimic Ni, then $m_1=8.66$ 
and $m_2=1.35$.~\cite{rf:Edwards2}\
These choices are based on the hole picture~\cite{rf:Kanamori}
in which 
0.6 $d$-hole per nickel atom 
occupies three sub-bands.
Moreover we assume the quadratic dispersion
for the magnon,~\cite{rf:Edwards2}
\begin{equation}
\omega^\prime_q=Dq^2,
\label{omegadq}
\end{equation}
where the stiffness constant is set such that
$Dk_1^2=0.2$ eV.
The density of states for the model (\ref{model}) is
shown in Fig.~\ref{fig:56}.

We investigate a model consisting of three equivalent 
sub-bands,
each of which has the dispersion $\varepsilon_k$ as given above.
The Hund's-rule coupling $J$
increases the mean-field exchange splitting
$g$ defined in (\ref{gmu}).
This effect, however,
is due to the longitudinal component of the coupling.
Therefore,
to investigate the transverse coupling effect, 
which has been discussed only little,
we take $g$ (or $g/n$) as a parameter 
for the one-particle bare interaction energy.
Thus, 
for given $g/n$, $J$ and $n$, a set of equations 
(\ref{eigeneqhj}) $\sim$ 
(\ref{tilderho})
are solved for $E_{k\mu\downarrow}$.
Results for a single-band model are obtained by
setting $J=0$ and $h_{\rm J}=0$.
Note that
the parameter $\overline{U}\equiv g/n$
represents 
$\overline{U}=U$ for the single-band model 
and $\overline{U}=U+J$ for the model with triply-degenerate orbitals. 
The parameters used below are $g/n$, $n$, $J$
and $\varepsilon_1$.

\subsection{Results}
\begin{figure}[tb]
 \parbox{\halftext}{
\epsfxsize = 6.1 cm
\centerline{\epsfbox{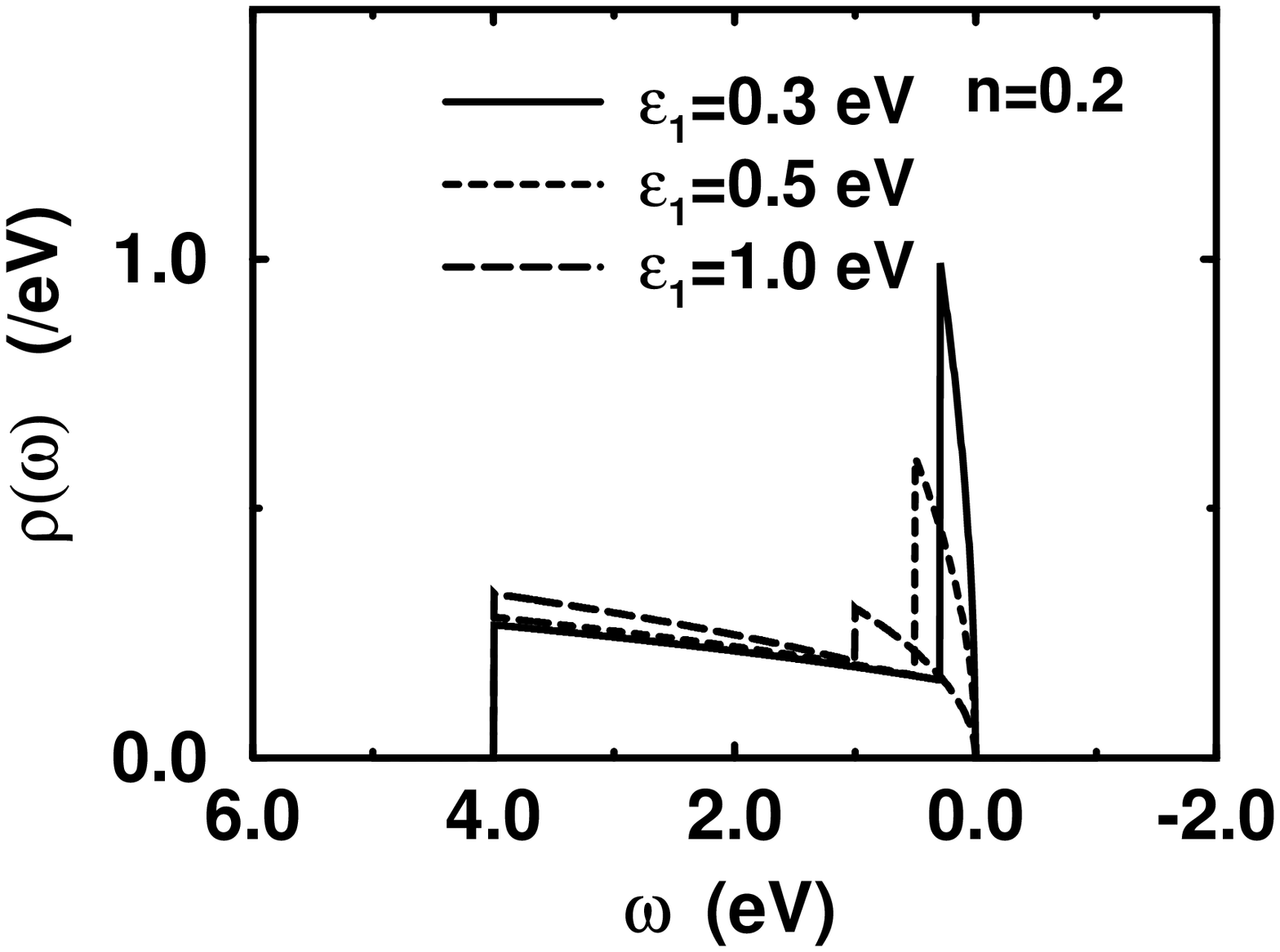}}
\caption{
Bare density of states of our model
for metallic nickel.
In the ferromagnetic state,
a carrier of total density $n=0.2$ per band
occupies up to the Fermi energy $\varepsilon_1$ (eV).
}
\label{fig:56}
}
 \hspace{5mm}
 \parbox{\halftext}{
\vspace{10mm}
\epsfxsize = 6.1 cm
\centerline{\epsfbox{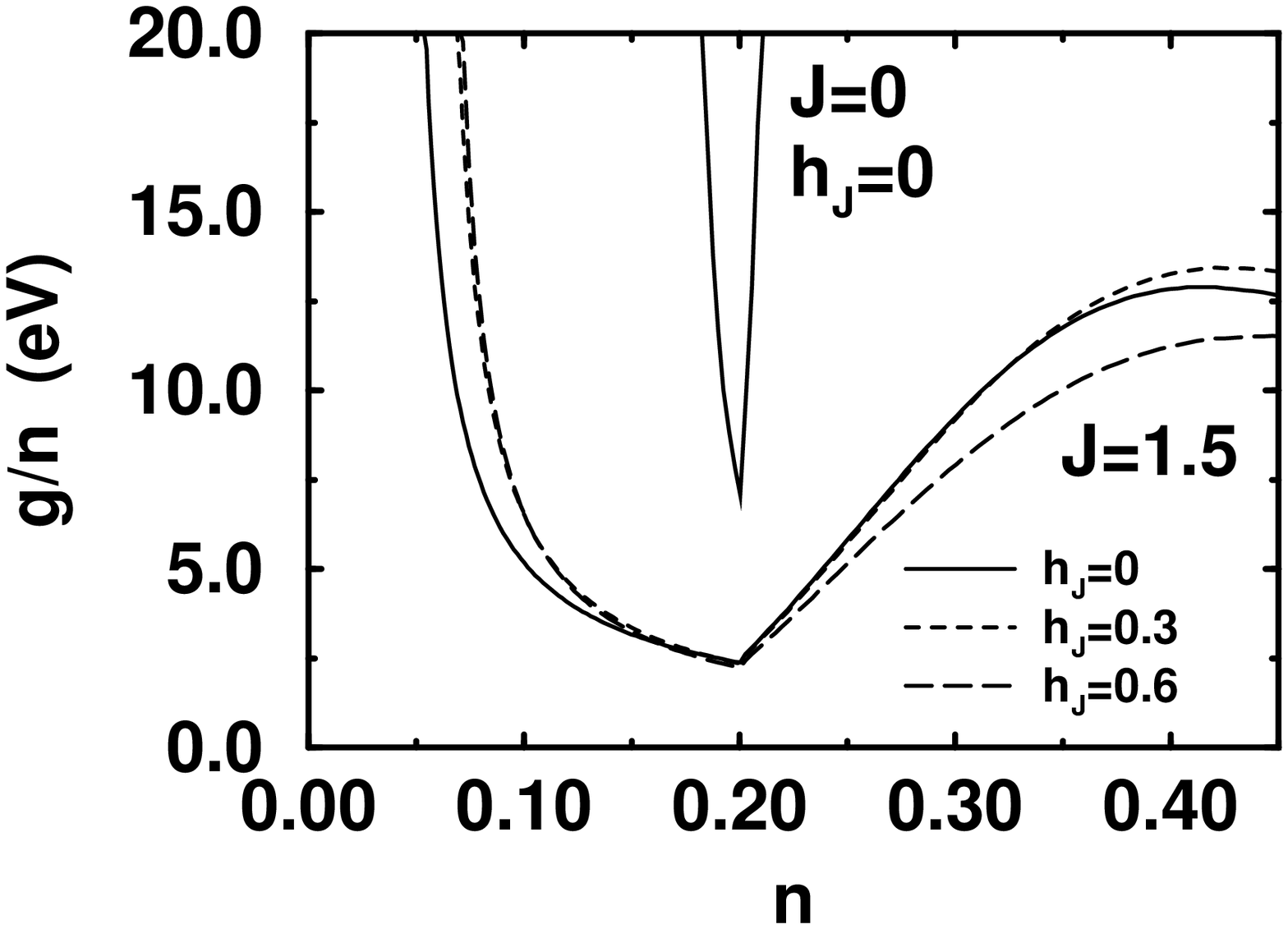}}
\caption{
The critical value of $g/n$ as a function of $n$ for 
$\varepsilon_{1}=0.3$ eV.
In the region below the curve,
the complete ferromagnetic state is unstable.
The curve 
for $J=0$ and $h_{\rm J}=0$ 
represents the result for the single-band model.
}
\label{fig:6}
}
\end{figure}

\begin{figure}[tb]
 \parbox{\halftext}{
\epsfxsize = 6.1 cm
\centerline{\epsfbox{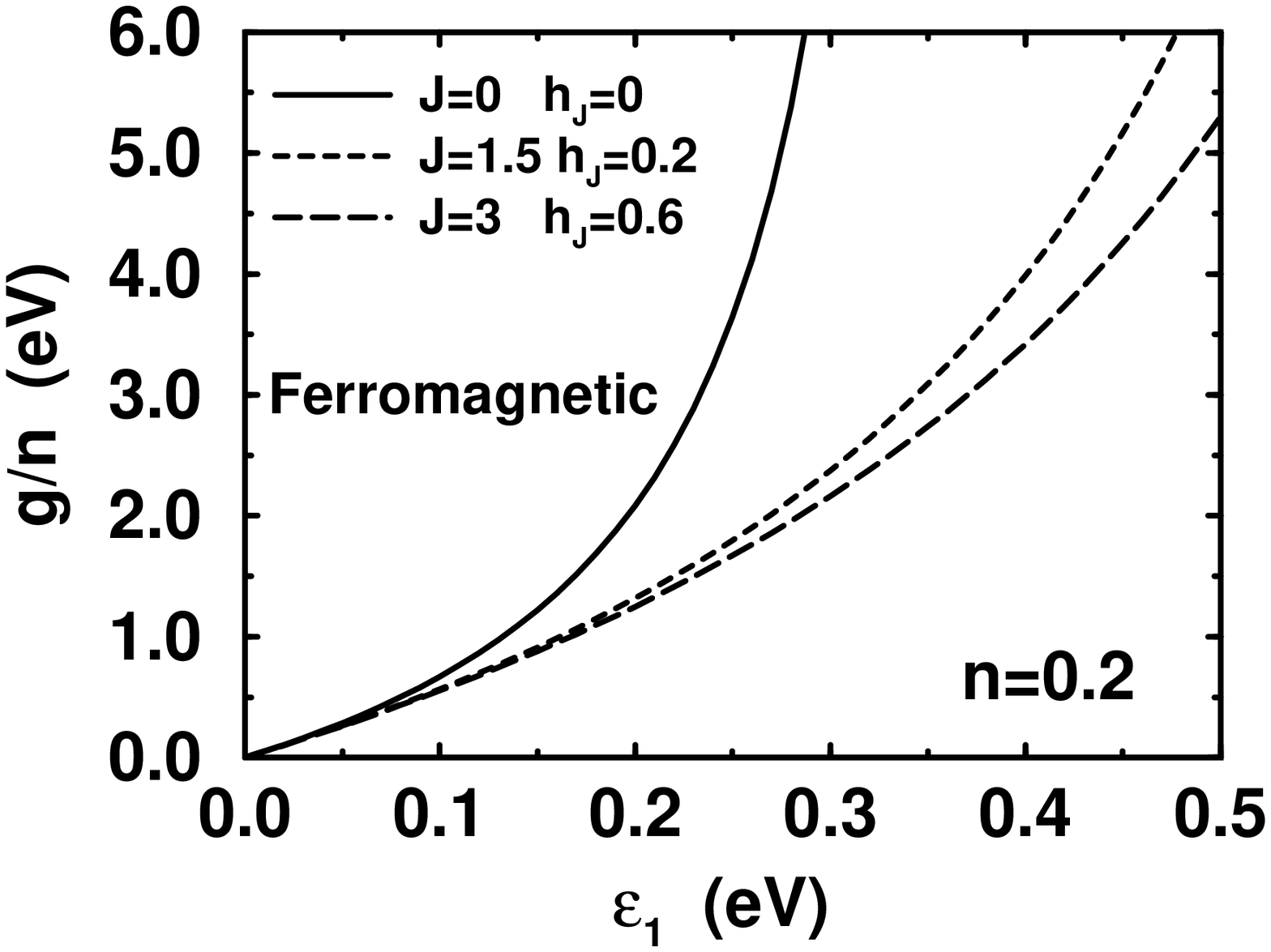}}
\caption{
The critical value of $g/n$ as a function of $\varepsilon_{1}$
for $n=0.2$.
In the region below the curve,
the complete ferromagnetic state is absolutely unstable.
The parameters $h_{\rm J}$ shown for $J=1.5$ and $J=3$
are those which give the most stringent criterion.
}
\label{fig:67}}
 \hspace{5mm}
 \parbox{\halftext}{
  \epsfxsize = 6 cm
\centerline{\epsfbox{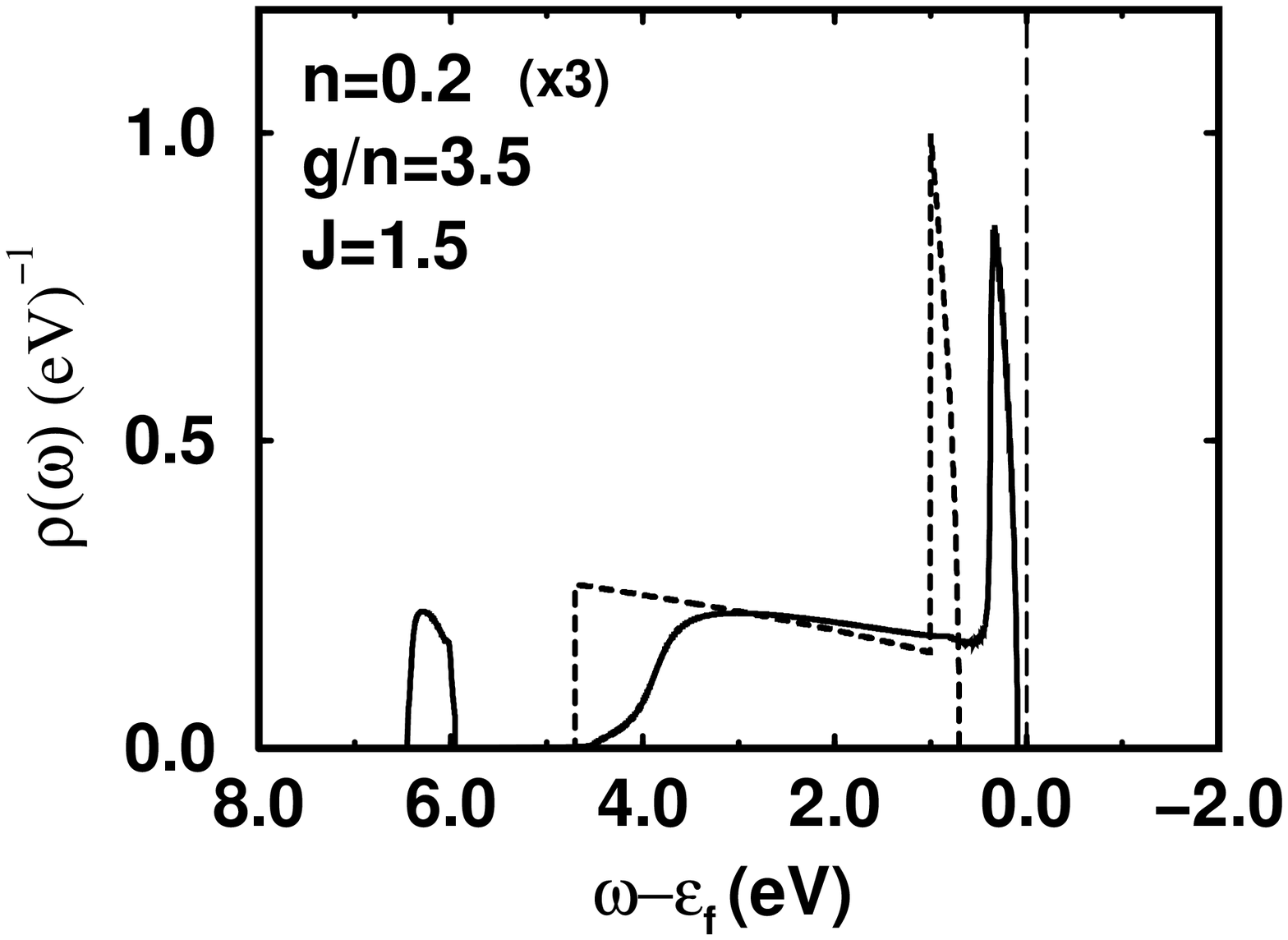}}
\caption{
Density of states $\rho(\omega)$
for  $g/n=3.5$, $J=1.5$, and $h_{\rm J}=0.2$.
The carrier density is $n=0.2$ for each of triply-degenerate bands.
The dashed curve represents the Hartree-Fock result.
The Fermi energy is $\varepsilon_{f}=\varepsilon_1=0.3$ eV.
}
\label{fig:7}}
\end{figure}

First,
we show the phase boundary determined by
$E_{k=0\mu\downarrow}=\varepsilon_{f}$
for the degenerate-band model,
as well as that of the single-band model
(the case $J=0$ and $h_{\rm J}=0$).
The critical interaction $\overline{U}_c\equiv g_c/n$ (eV) 
as a function of the carrier density $n$
is shown in Fig.~\ref{fig:6} for $\varepsilon_{1}=0.3$ eV.
Since we should have $g/n\LA 4$ eV and $n=0.2$ for Ni,
we conclude that the Hund's-rule coupling
is necessary to stabilize the ferromagnetic ground state.
Here we give some examples of the results:
For $n=0.2$ and $g/n=4$ eV, 
the ferromagnetic state becomes 
unstable for $J\LA0.3$ eV, and 
$h_{\rm J}$ becomes non-zero for $J\GA1.3$ eV,
while $h_{\rm J}=0.7$ for $J=3$ eV.

After the importance of the the Hund's-rule coupling
is shown,
the next question is that concerning a peculiarity in the 
bare density of states.
We know that a large density of states is not necessary
for ferromagnetism in the ferromagnetic Kondo lattice model.
However, in metallic Ni, the carrier density is so low that
it is not clear if the Hund's-rule coupling alone 
can stabilize the ferromagnetic state.
To see the effect,
one may change the parameter $\varepsilon_{1}$,
which is set above as 0.3 eV.
We display the phase boundary $g/n$ as a function of 
$\varepsilon_{1}$ for $n=0.2$
(Fig.~\ref{fig:67}).\
Note that
states of total weight $n=0.2$
occupy up to $\varepsilon_{1}$ (eV) within 
the band of 4\,eV wide.
Thus a small $\varepsilon_{1}$ represents a strong peculiarity,
favoring ferromagnetism.
(See Fig.~\ref{fig:56}.)
We see in Fig.~\ref{fig:67} that
the Hund's-rule coupling is necessary to 
keep the relevant region 
around $(\varepsilon_{1},g/n)\simeq (0.3,4.0)$ eV ferromagnetic.
Still, a large bare density-of-states is also required,
as concluded in Ref.~\citen{rf:Okabe1}.
Note that we require
$\varepsilon_{1}$
to be as large as
$4(k_1/k_{\rm D})^2=1.37$ eV
to give
a simple parabolic band, for which $m_1=m_2$.
Even with $\varepsilon_{1}\sim 0.5$ eV, 
the peak structure in the density of states is pronounced,
as shown in Fig.~\ref{fig:56}.

Although the phase boundary in Fig.~\ref{fig:6} 
around $n\simeq 0.2$ does not depend 
strongly on $h_{\rm J}$,
the coupling in terms of $h_{\rm J}$ plays 
an important role.
This point is 
clearly seen in the renormalized density-of-states curve $\rho(\omega)$.
To investigate whether we obtain the satellite structure,
we calculate $\rho(\omega)$ defined by
\begin{equation}
\rho(\omega)\equiv
-\frac{1}{\pi}{\rm Im}
\frac{1}{\Lambda}\sum_k
\frac{1}{\omega-(\varepsilon_k-\varepsilon_{f})-
\Sigma(k,\omega)},
\label{dosrho}
\end{equation}
where $\Sigma(k,\omega)$ is given by (\ref{eigeneqhj2}).
In order to estimate the position of a satellite structure,
one may neglect the momentum dependence of $\Sigma(k,\omega)$,
or $\omega_q$ may be replaced with the quantity averaged over $q$:
$\overline{\omega}_q\equiv \frac{3}{5}Dk_{\rm D}^2$.~\cite{rf:Okabe2}\
Then, the calculation  is straightforward, and 
the result is shown in Fig.~\ref{fig:7}
for $\varepsilon_{1}=0.3$ eV.

It is important to note that 
we can assume parameters so as to reproduce the 6 eV
satellite structure while preventing the instability of 
the ferromagnetic state.
In the figure, 
we set
$g/n=3.5$ eV, $J=1.5$ eV and $n=0.2$.
The parameter 
$h_{\rm J}=0.2$ is determined 
to minimize $E_{k=0\mu\downarrow}$.
These values for $g/n$ and $J$ are
physically reasonable for metallic nickel.
In the calculation with the other set of parameters,
we observed that the position of the satellite depends on $h_{\rm J}$,
if it is regarded formally as a free parameter.
For example, for the same set of parameters
except for $h_{\rm J}$,
the satellite appears at 5 eV and 7.7 eV
for $h_{\rm J}=0$ and 1, respectively.
The weight of the satellite also depends on $h_{\rm J}$
and is an increasing function of $h_{\rm J}$.
This is physically interpreted as follows:
The weight in the low-lying part of spectrum
(the part other than that due to the satellite) decreases 
as the local constraint from the coupling $J$ becomes effective,
or as $h_{\rm J}$ increases.~\cite{rf:Okabe2}\

\section{Conclusion}
\label{DC}
We investigated the effect of the Hund's-rule 
coupling and the orbital degeneracy on the basis of the variational
treatment
of a general model for the itinerant ferromagnet.
Our model includes the ferromagnetic Kondo lattice model and 
the Hubbard model as special cases.
It was also applied
to the ferromagnetism in Ni.
To investigate instability due to individual particle excitation,
we took into account
the multiple magnon scattering effect
and reproduced some of the results which were
derived in the Hubbard model by many authors.
Generally, in the theory of itinerant ferromagnetism,
it is  important to meet the requirement imposed by 
the rotational symmetry of the model Hamiltonian.
In terms of the diagram technique,
one should take into account the vertex correction
in accordance with the Ward-Takahashi 
identity.~\cite{rf:Brandt,rf:Hertz,rf:Matsumoto}\
In this respect, 
we avoided complexity
by making use of proper variational states.
We derived the required expressions  variationally
without any ambiguous assumption.
Variational derivation is advantageous
because we do not have to be concerned with 
the problem of selecting the type of diagrams to be summed over
and we are free from 
a convergence problem of the summed infinite series.
Moreover,
the result for the instability has an exact significance,
since it follows the variational principle.

We clarified the physics introduced by the 
Hund's-rule coupling by treating a two-band model
and a model simulating metallic nickel.
For the former,
we noted that attention should be paid to
the treatment of the low-density limit
of the degenerate-band model,
owing to the singular behavior
of the spin-wave stiffness constant $D$.
We displayed figures for $D$  which show that
the correction to the result of the RPA becomes smaller 
when the spontaneous magnetization becomes larger.
Investigation of the individual-particle excitation 
gives a more stringent condition 
for the instability of the ferromagnetic state
than that given by the spin-wave instability.
However, in both cases, we observed a common tendency
that a two-band model is quite stable 
compared with a single-band model.
As a function of the Hund's-rule coupling $J$,
we investigated how 
two bands are ferromagnetically coupled.
In the case where one of them constitutes localized spins,
$J$ is quite effective, as expected.
In the other case, where
the two bands are equivalent,
$J$ stabilizes the ferromagnetic state
in the regime $J\LA W$.
We conclude for Ni that 
a peculiarity of the density-of-state 
curve at the Fermi level is 
indispensable, while the effect of 
Hund's-rule coupling is also required.
In this respect,
ferromagnetism in Ni differs from the 
double exchange ferromagnet,
in which the complete ferromagnetic state is made stable
only by the Hund's rule coupling. 
However, it is conceptually simple
to discuss itinerant ferromagnetism 
on the basis of a unified model as we proposed in this article.
It was also shown that the effect of the Hund's-rule coupling 
including its transverse component 
can explain the 6 eV satellite structure as well as
the ferromagnetic ground state of metallic nickel.

\section*{Acknowledgments}
The author would like to thank Professor K. Yamada for discussions
and T. Ichinomiya and S. Yoda for their helpful 
assistance in preparing the manuscript and figures.
This work is supported  by
Research Fellowships of the Japan Society for the
Promotion of Science for Young Scientists.


\begin{thebibliography}{99}

\bibitem{rf:Kanamori}J. Kanamori,
\PTP{30,1963,275}.
\bibitem{rf:Gutzwiller}M. C. Gutzwiller,
\PRL{10,1963,159}.
\bibitem{rf:Hubbard}J. Hubbard,
Proc. R. Soc. London {\bf A276} (1963), 238.
\bibitem{rf:Nagaoka}Y. Nagaoka,
\PR{147,1966,392}.
\bibitem{rf:Tasaki}H. Tasaki,
\PRL{69,1992,1608}.\\
A. Mielke and H. Tasaki,
Commun. Math. Phys. {\bf 158} (1993), 341.
\bibitem{rf:Penc}K. Penc, H. Shiba, F. Mila and T. Tsukagoshi,
\PR{B54,1996,4056}.
\bibitem{rf:comment0}
In the strong-coupling regime of the Hubbard model,
it is generally more difficult to prove 
instability of the ferromagnetic state
for the density just around half-filling
than for low densities.
This is concluded at least by a variational treatment,
e.g.,  in Fig.~\ref{fig:3} of this article.
Although this does not necessarily mean that 
the ferromagnetic state is actually realized  
in the single-band Hubbard model,
this conclusion is consistent with 
the result of Nagaoka.~\cite{rf:Shastry}\
Therefore, we simply call this hypothetical ferromagnetic state 
the Nagaoka ferromagnetic state,
although the origin of its stability 
may not be regarded as due to a coherent motion of 
doped holes, as first envisaged by Nagaoka.





\bibitem{rf:Antonides}E. Antonides, E. C. Janse and G. A. Sawatzky,
\PR{B15,1977,1669}.
\bibitem{rf:Fuggle}J. C. Fuggle, P. Bennett, 
F. U. Hillebrecht, A. Lenselink and G. A. Sawatzky,
\PRL{49,1982,1787}.
\bibitem{rf:Bennett}P. Bennett, J. C. Fuggle,
F. U. Hillebrecht, A. Lenselink and G. A. Sawatzky,
\PR{B27,1983,2194}.
\bibitem{rf:XPS1}S. H\"ufner and G. K. Wertheim,
Phys. Lett. {\bf 51A} (1975),299.
\bibitem{rf:XPS2}P. C. Kemeny and N. J. Shevchik,
Solid State Commun. {\bf 17} (1975), 255.
\bibitem{rf:UPS1}R. J. Smith, J. Anderson, J. Hermanson
and G. J. Lapeyre,
Solid State Commun. {\bf 21} (1977), 459.
\bibitem{rf:UPS2}C. Guillot, Y. Ballu, J. Paign\'e, J. Lecante,
K. P. Jain, P. Thiry, R. Pinchaux, Y. P\'etroff and L. M. Falicov,
\PRL{39,1977,1632}.
\bibitem{rf:Penn}D. Penn,
\PRL{42,1979,921}.
\bibitem{rf:Treglia}G. Treglia, F. Ducastelle and D. Spanjaard,
J. Physique, {\bf 41} (1980), 281.
\bibitem{rf:Edwards1}
D. M. Edwards,
\JL{J. Appl. Phys.,39,1968,481}.
\bibitem{rf:Brandt}U. Brandt,
Z. Phys. {\bf 244} (1971), 217.
\bibitem{rf:Roth}L. M. Roth,
\PR{186,1969,428}.
\bibitem{rf:Hertz}J. A. Hertz and D. M. Edwards,
\JP{F3,1973,2174}.
\bibitem{rf:Edwards2}D. M. Edwards and J. A. Hertz,
\JP{F3,1973,2191}.
\bibitem{rf:Liebsch1}A. Liebsch,
\PRL{43,1979,1431}.
\bibitem{rf:Liebsch2}A. Liebsch,
\PR{B23,1981,5203}.
\bibitem{rf:Igarashi1}
J. Igarashi,
{\it Electron Correlation and Magnetism in Narrow-Band Systems},
ed. T. Moriya (Springer Verlag, 1981), p168.


\bibitem{rf:Okabe1}T. Okabe,
\JPSJ{65,1994,1056}; \andvol{66,1997,No.7}.

\bibitem{rf:Okabe2}T. Okabe,
\PTP{97,1997,21}; \andvol{97,1997,559}.

\bibitem{rf:Morruzi}V. L. Morruzi, J. F. Janak and A. R. Williams,
{\it Calculated Electronic Properties of Metals}
(Pergamon, 1978).
\bibitem{rf:comment}
In view of experimental results,~\cite{rf:Fuggle,rf:Bennett}
this is not an unreasonable assumption.


\bibitem{rf:Roth1}L. M. Roth,
J. Phys. Chem. Solids {\bf 28} (1967), 1549;
J. Appl. Phys.{\bf 39} (1968), 474.
\bibitem{rf:Shastry}B. S. Shastry, H. R. Krishnamurthy and P. W. Anderson,
\PR{B41,1990,2375}.




\bibitem{rf:Matsumoto}H. Matsumoto, H. Umezawa, S. Seki
and M. Tachiki,
\PR{B17,1978,2276}.



\end{thebibliography}
\end{document}